\documentclass[pra, aps, twocolumn, groupedaddress, showpacs, superscriptaddress]{revtex4}

\usepackage{amssymb,amsmath,amsthm,color,graphicx,times,graphicx}
\usepackage{hyperref}
\usepackage{subfigure}
\usepackage{times}
\usepackage{bbold}
\usepackage{color}

\theoremstyle{plain}

\theoremstyle{definition}

\begin{document}
\title{Influence of Stark-shift on quantum coherence and non-classical correlations for two two-level atoms interacting with a single-mode cavity field}
\author{Abdallah Slaoui}\email{abdallah\_slaoui@um5.ac.ma}\affiliation{LPHE-Modeling and Simulation, Faculty of Sciences, Mohammed V University in Rabat, Morocco.}
\author{Ahmed Salah}\email{asalah3020@gmail.com}\affiliation{Mathematics and Theoretical Physics Department, Nuclear Research Center (NRC), Atomic Energy Authority, Cairo, 13759, Egypt.}\affiliation{Abdus Salam International Centre for Theoretical Physics (ICTP), Strada Costiera, 11 I-34151, Trieste, Italy.}
\author{Mohammed Daoud}\email{m\_daoud@hotmail.com}\affiliation{Abdus Salam International Centre for Theoretical Physics (ICTP), Strada Costiera, 11 I-34151, Trieste, Italy.}\affiliation{Department of Physics, Faculty of Sciences, University Ibn Tofail, Kenitra, Morocco.}

\begin{abstract}
An exact analytic solution for two two-level atoms coupled with a multi-photon single-mode electromagnetic cavity field in the presence of the Stark shift is derived. We assume that the field is initially prepared in a coherent state and the two atoms are initially prepared in excited state. Considering the atomic level shifts generated by the Stark shift effect, the dynamical behavior of both quantum coherence (QC) measured using a quantum Jensen-Shannon divergence and of quantum correlations captured by quantum discord (QD) are investigated. It is ﻿shown that the intensity-dependent Stark-shift in the cavity and the number of coherent state photons plays a key role in enhancing or destroying both QC and QD during the process of intrinsic decoherence. We remarked that increasing the Stark-shift parameters, the frequencies of the transition for the mode of the cavity field, and photons number destroy both the amount of QC and QD and affected their periodicity. More importantly, QC and QD exhibit similar behavior and both show a revival phenomenon. We believe that the present work shows that the quantum information protocols based on physical resources in optical systems could be controlled by adjusting the Stark-shift parameters.

\par

\end{abstract}
\date{\today}

\maketitle

\section{Introduction}
Over the past decades, the two-levels atoms constitute valuable tools for both theoretical and experimental investigations in various fields of modern physics such as collision physics \cite{Shore1990,Scully1997} and quantum optics \cite{Allen1997,Li2019}. They were considered to be the simplest quantum model used to solve many problems of light-matter interaction and for the fundamental building of modern applications in quantum control \cite{Nakamura1999} and quantum processing \cite{Treutlein2014}. Further, there are many physical systems recognized as a possible candidate for quantum processing such as the cavity QED which studies the interaction between the individual atoms and a single-mode electromagnetic field inside a cavity \cite{Lukin2011}, the artificial two-level atoms based on the superconducting qubits \cite{Kang2016,Khan2018}, ion traps \cite{Steane1997} and quantum dots \cite{Michler2017}.\par

On the other hand, the Dicke model \cite{Dicke1954}, describes the interaction between a collection of $N$ two-level atoms and a quantized electromagnetic field, has been extensively used in quantum optics. It gives an opportunities to simulate quantum optical and condensed matter phenomena. This class of models has been employed to investigate quantum phase transition with a coupled optical cavity \cite{Abdel-Rady20172}. Also, a new physical phenomenon of the splitting of the dressed states is explained as a manifestation of an $n$-photon coupling between them, i.e., it represents an $n$-photon ac Stark effect parameter\cite{Rudolph1998}. Moreover, a three-level $\Lambda$-type atom interacting with a two-mode of electromagnetic cavity field surrounded by a nonlinear Kerr-like medium with decay rates have been considered in \cite{Farouk2017}. The time-dependent interaction between the time-dependent field and a two-level atom with one mode electromagnetic field have been explored in \cite{Abdel-Wahab2019}. In addition, the time-dependent fields were also used for controlling quantum states in several protocols and shown various useful applications, especially in carrying out rapid population transfers by the invariant-based inverse engineering method in a three-level system \cite{Chen2012}, realizing the perfect adiabatic state transfer protocols via an intermediate lossy system \cite{Baksic2016,Chen2017} and performing fast population transfer for ground states in multiparticle cavity QED systems via the adiabatic passage \cite{Chen2014,Chen2015}.

\par
Quantum entanglement (QE), as a special type of quantum correlations, is a valuable resource for several tasks of quantum information processing and has a crucial role in setting the boundary between quantum and classical worlds \cite{Nielsen2000,Horodecki2009}. It has recently attracted significant attention as an essential ingredient for powerful applications including quantum teleportation \cite{Bennett1993}, dense coding \cite{Bennett1992}, quantum cryptography \cite{Jennewein2000} and telecloning \cite{Murao1999}. In this regard, various proper entanglement measures have been proposed by employing a rigorous mathematical framework \cite{Vedral1997} and their time evolution has been studied in several quantum systems of two-level atoms \cite{Braun2002,Slaoui20182,Yu2004}. It was believed that QE is tightly related to quantum correlations. However, according to various studies \cite{Datta2005,Datta2007}, it has been proved experimentally that there exist some separable quantum states that are very useful in practical quantum technology and that QE is not the only meaningful type of quantum correlations. For instance, quantum speedup with separable states \cite{Braunstein1999,Meyer2000}, quantum nonlocality without QE \cite{Bennett1999,Niset2006} and the efficiency of deterministic﻿ quantum computation with one qubit \cite{Lanyon2008}. In this sense, considerable efforts have been devoted over the last years to characterize and to quantify quantum correlations that may be contained in a multipartite system.\par

The problem of quantifying non-classical correlations beyond QE in a generic quantum state hasn't solved yet and their investigation in closed and open multipartite systems is one of the most challenging topics in the literature. Historically, entropic quantum discord (QD) is an alternative approach to investigate quantum correlations for an arbitrary bipartite state even those which are separable. This measure was first proposed independently by Ollivier et al., \cite{Ollivier2001} and Henderson et al., \cite{Henderson2001}. They have shown that QD is more general and goes beyond QE and it coincides with the entanglement of formation for pure bipartite states. In addition to that, the QD obeys a conservation law with the QE based upon a monogamic principle of the quantum correlation distribution \cite{Fanchini2011} and it is related to the QE irreversibility \cite{Cornelio2011}. Interestingly, Dakić et al.,\cite{Dakic2012} showed that a correlated separable state (i.e., the separable state with non-zero QD) can outperform entangled states and it exhibits better performance for remote quantum state preparation. Datta et al.,\cite{Datta2008} confirmed that QD is a better figure of merit for providing computational speedup compared with classical states in quantum computation models. Moreover, Werlang et al.,\cite{Werlang2009} showed that QD is more robust than QE against decoherence in Markovian environments. It is interesting to stress that other discord-like measures have been introduced such as one-way quantum deficit \cite{Horodecki2005}, local quantum uncertainty \cite{Girolami2013,Slaoui2018,Slaoui2019}, local quantum Fisher information \cite{Kim2018,Slaoui20192} and quantum interferometric power \cite{Girolami2014}.\par

On the other side, one of the main challenges in the practical implementation of quantum computing and in the physical realization of quantum technologies is how to protect quantum coherence (QC) and overcome decoherence \cite{Zurek2003}. This is due to the realistic quantum systems are naturally open and very brittle, which leads to loss of information and the intrinsic physical quantum properties from the system to the environment. So, it is important to study both QD and QC dynamics in open quantum systems when loses their coherence. QC also provides essential power for quantum information processing and a theoretic framework for their quantification in quantum states has been developed \cite{Aberg2006,Aberg2014,Baumgratz2014}. The new fundamental and challenging task in this field, from both theoretical and experimental perspectives, consists of how to characterize and quantify the interplay between these different kinds of quantum correlation. In this sense, the role of QC to characterize QE and quantum thermodynamics was discussed in \cite{Streltsov2016} and in \cite{Santos2019,Narasimhachar2015} respectively. \par

Inspired by these works, we would study the QD and QC dynamics in an open quantum system consisting of two two-level atoms that are connected by a single-mode electromagnetic cavity field in the presence of the Stark-shift effect. We are especially interested in how intensity-dependent Stark-shift in the cavity and the number of coherent state photons would affect both QD and QC behaviors in this model. This paper is structured as follows: In sec.\ref{sec:Model}, we firstly introduce the considered physical model and then we find an exact solution of the density matrix in the Schrödinger picture when these two atoms are initially prepared in their excited state. In sec.\ref{sec:QC}, we give the explicit expression of QC by employing the concepts of the quantum Jensen-Shannon divergence. The time evolution behavior of the QC as a function of the parameters of the considered system and its environment is also discussed. Sec.\ref{sec:QD} is devoted to the effect of Stark-shift in the cavity on﻿ the temporal evolution of correlations captured by QD. A brief summary of the main obtained results closes this paper.

\begin{widetext}
	
	\begin{figure}[h]
	\begin{minipage}[b]{.50\linewidth}
		\centering
		\vfill $\left(a\right)$
		\includegraphics[scale=0.3]{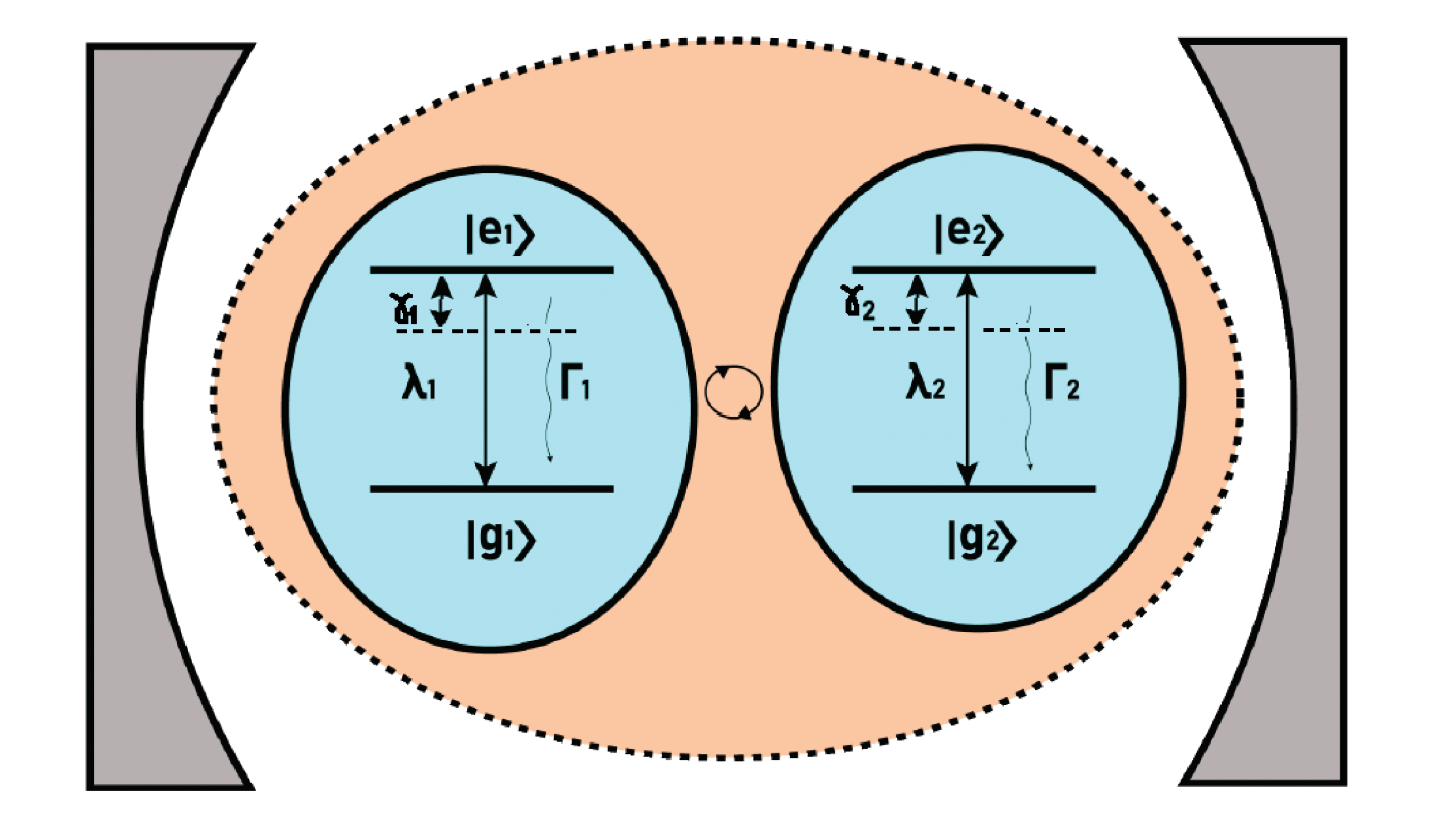}
	\end{minipage}
	\begin{minipage}[b]{.45\linewidth}
		\centering
		\vfill $\left(b\right)$
		
		\includegraphics[scale=0.4]{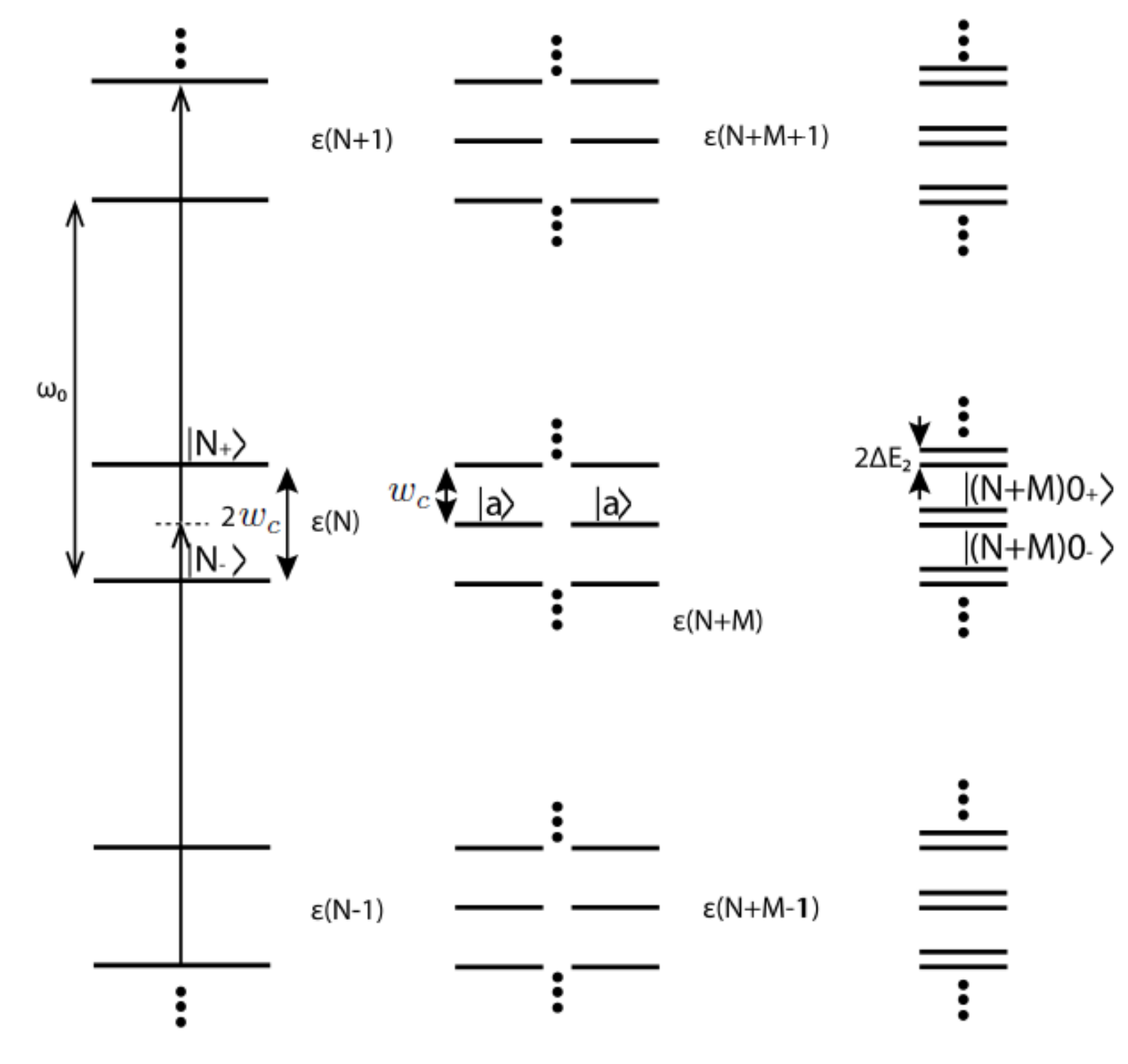}
	\end{minipage}
	\caption{(a) Energy level diagram for two two-level atoms interacting to a single-mode cavity field (b) graphical representation of the dynamic Stark splitting of the effective two two-level atoms.
}\label{fig11}	
	\end{figure}

\end{widetext}
\section{The Model: Hamiltonian and Solution}\label{sec:Model}
In this work, the system under consideration consists of two 2-level atoms (considered as two qubits $A$ and $B$) having ground states $\left| {{g_i}} \right\rangle$ and excited states $\left| {{e_i}} \right\rangle$ ($i=1,2$) which are coupled with a single-mode electromagnetic cavity field. In our consideration, we assume that the interaction occurs in the presence of the Stark-shift effect as shown schematically in Fig.(\ref{fig11}). This Stark effect induces the splitting and shifting of spectral energy levels due to the presence of an electromagnetic cavity field \cite{Carollo2003,Salah2018,Abdel-Rady2017}. In this picture, the total Hamiltonian of the system is given by ($\hbar  = 1$)
\begin{equation}
{\hat H }= {\hat H_{\rm A}} + {\hat H_{\rm F}} + {\hat H_{\rm A-F}} + {\hat H_{\rm Stark}},
\end{equation}
where ${\hat H_{\rm A}}$ and ${\hat H_{\rm F}}$ describes respectively the two atoms Hamiltonian and the quantized electromagnetic cavity field Hamiltonian, which can be written as
\begin{equation}
{\hat H_{\rm A}} + {\hat H_{\rm F}} = \frac{1}{2}\sum\limits_{j = 1}^2 {{w_z}\hat S_z^j}  + {w_c}{{\hat a}^\dag }\hat a,
\end{equation}
where $\hat S_z^j = \left| {{e_j}} \right\rangle \left\langle {{e_j}} \right| - \left| {{g_j}} \right\rangle \left\langle {{g_j}} \right|$ is the energy operator of $i$th atom, ${w_z}$ and ${w_c}$ denotes the transition frequency and the frequency of the mode of the cavity field, respectively and  ${\hat a}^\dag$ (${\hat a}$) is the creation (annihilation) operators of the cavity field which they satisfy the usual bosonic commutation relations. The interaction Hamiltonian ${\hat H_{\rm A-F}}$ between the atoms and the electromagnetic cavity field is given by
\begin{equation}
{\hat H_{\rm A-F}} = \sum\limits_{j = 1}^2 {{\lambda _j}\left( {\hat a\hat S_ + ^j + {{\hat a}^ \dag }\hat S_ - ^j} \right)},
\end{equation}
where $\lambda _j$ ($j=1,2$) is the coupling strength between the atom $j$ and the mode of the cavity field and $\hat S_ + ^j = \left| {{e_j}} \right\rangle \left\langle {{g_j}} \right|$ and $\hat S_ - ^j = \left| {{g_j}} \right\rangle \left\langle {{e_j}}\right|$ represents the dipole raising and lowering operators, respectively. The Hamiltonian describing the effect of Stark shift in the cavity can be written as \cite{Salah2018}
\begin{equation}
{\hat H_{\rm Stark}} = \left( {{\gamma _1} + {\gamma _2}} \right)\sum\limits_{j = 1}^2 {\hat S_z^j}  + \frac{1}{2}\left( {{\gamma _1} + {\gamma _2}} \right){{\hat a}^\dag }\hat a\sum\limits_{j = 1}^2 {\hat S_z^j},
\end{equation}
where ${\gamma _1}$ and ${\gamma _2}$ describe the intensity-dependent Stark-shift in the cavity. The dynamic Stark shift parameters and effect has been included in two-photon absorption studies in the past. The Stark effect to first order in a perturbative calculation in two-photon equal frequency absorption has been studied \cite{Alsing1987}. Furthermore, the importance of the Stark-shift terms in an exact calculation of two-photon equal-frequency absorption has been demonstrated \cite{Puri1988}. Moreover, Ashraf and Zubairy \cite{Ashraf1990} included this power-dependent effect in their study of the equal-frequency two-photon micromaser. Also, Gou \cite{Gou1990} discussed how to eliminate the Stark shifts through the use of correlated two-mode field states in unequal-frequency absorption. All of these studies investigated the case of exact two-photon resonance. The general expression of the time-dependent wave function of an atom $j$ ($j=1,2$) in the system under consideration can be written as 
\begin{equation}
\left| {{\psi _n^{j}}\left( t \right)} \right\rangle  = {A_j}\left( t \right)\left| {{e_j},n} \right\rangle  + {B_j}\left( t \right)\left| {{g_j},n + 1} \right\rangle,
\end{equation}
with ${A_j}\left( t \right)$ and ${B_j}\left( t \right)$ are the probability amplitudes. Consequently, after interacting with a single-mode electromagnetic cavity field, the state vector of the whole system $\left| {{\psi}\left( t \right)} \right\rangle$ takes the following form
\begin{eqnarray}
\left| {{\psi}\left( t \right)} \right\rangle  &=& \sum\limits_{n = 0}^\infty\left[A_{n}\left( t \right)\left| {{e_1},{e_2},n} \right\rangle  + C_{n+1}\left( t \right)\left| {{e_1},{g_2},n + 1} \right\rangle \right.\label{state} \\
	&~&\left.+D_{n+1}\left( t \right)\left| {{g_1},{e_2},n + 1} \right\rangle  + B_{n+2}\left( t \right)\left| {{g_1},{g_2},n + 2} \right\rangle \right] \nonumber 
\end{eqnarray}
with $A_{n}\left( t \right)={A_1\left( t \right)}{A_2\left( t \right)}$, $C_{n+1}\left( t \right)={A_1\left( t \right)}{B_2\left( t \right)}$, $D_{n+1}\left( t \right)={B_1\left( t \right)}{A_2\left( t \right)}$ and $B_{n+2}\left(t\right)={B_1\left(t\right)}{B_2\left(t\right)}$. To determine the probability amplitudes of the state (\ref{state}), it is necessary to solve the time-dependent Schrödinger equation
\begin{equation}
i \frac{{\partial \left| {{\psi}\left( t \right)} \right\rangle }}{{\partial t}} = {\hat H}\left| {{\psi }\left( t \right)} \right\rangle.
\end{equation}
After some simplifications and by using the following creation and annihilation operators actions
\begin{equation}
	\begin{array}{l}
		a\left| n \right\rangle  = \sqrt n \left| {n - 1} \right\rangle, \\
		{a^\dag }\left| n \right\rangle  = \sqrt {n + 1} \left| {n + 1} \right\rangle,
	\end{array}
\end{equation}
we obtain the following coupled differential equations
\begin{align}
i\frac{{\partial {A_n}\left( t \right)}}{{\partial t}} &= {\xi _1}{A_n}\left( t \right) + {\lambda _2}\sqrt {n + 1} {C_{n + 1}}\left( t \right) \notag\\&+ {\lambda _1}\sqrt {n + 1} {D_{n + 1}}\left( t \right), \label{An}
\end{align}
\begin{align}
i\frac{{\partial {C_{n + 1}}\left( t \right)}}{{\partial t}} &= {\lambda _2}\sqrt {n + 1} {A_n}\left( t \right) + {\xi _2}{C_{n + 1}}\left( t \right) \notag\\&+ {\lambda _1}\sqrt {n + 2} {B_{n + 2}}\left( t \right), \label{Cn}
\end{align}
\begin{align}
i\frac{{\partial {D_{n + 1}}\left( t \right)}}{{\partial t}} &= {\lambda _1}\sqrt {n + 1} {A_n}\left( t \right) + {\xi _3}{D_{n + 1}}\left( t \right) \notag\\&+ {\lambda _2}\sqrt {n + 2} {B_{n + 2}}\left( t \right),\label{Dn}
\end{align}
and
\begin{align}
i\frac{{\partial {B_{n + 2}}\left( t \right)}}{{\partial t}} &= {\lambda _1}\sqrt {n + 2} {C_{n + 1}}\left( t \right) + {\lambda _2}\sqrt {n + 2} {D_{n + 1}}\left( t \right) \notag\\&+ {\xi _4}{B_{n + 2}}\left( t \right).\label{Bn}
\end{align}
The parameters appearing in these differential equations are
\begin{equation}
{\xi _1} = {w_z} + {w_c}n + \left( {{\gamma _1} + {\gamma _2}} \right)\left( {2 + n} \right),
\end{equation}
\begin{equation}
{\xi _2} = {\xi _3} = {w_c}\left( {n + 1} \right),
\end{equation}
and
\begin{equation}
{\xi _4} =  - {w_z} + {w_c}\left( {n + 2} \right) - \left( {{\gamma _1} + {\gamma _2}} \right)\left( {n + 4} \right).
\end{equation}
We assume that two atoms are initially prepared in their excited states $\left| e_1 \right\rangle$ and $\left| e_2 \right\rangle$. We also assume that the cavity electromagnetic field is prepared in the Glauber coherent states
\begin{equation}
\left| \alpha \right\rangle=\exp\left( -\frac{|\alpha|^{2}}{2}\right)\sum\limits_{n = 0}^\infty \frac{\alpha^{n}}{\sqrt {n!}} \left| n\right\rangle
\end{equation}
where $\alpha\in\mathbb{C}$. In this scheme, the initial state is given by
\begin{equation}
\left| {\psi \left( 0 \right)} \right\rangle  = \left| {{e_1},{e_2},\alpha} \right\rangle.
\end{equation}
Comparing with the equation (\ref{state}), the initial probability amplitudes are
\begin{equation}
{A_n}\left( 0 \right) = \exp\left( -\frac{|\alpha|^{2}}{2}\right)\frac{\alpha^{n}}{\sqrt {n!}},
\end{equation}
and
\begin{equation}
{C_{n + 1}}\left( 0 \right) = {D_{n + 1}}\left( 0 \right)= {B_{n + 2}}\left( 0 \right) = 0.
\end{equation}
To solve the differential equations (\ref{An}), (\ref{Cn}), (\ref{Dn}) and (\ref{Bn}), we set
\begin{equation}
A_{n}\left( t\right)=a_{n}\left( t\right)\frac{\alpha^{n}}{\sqrt {n!}},
\end{equation}
\begin{equation}
C_{n+1}\left( t\right)=c_{n+1}\left( t\right)\frac{\alpha^{n}}{\sqrt {n!}},
\end{equation}
\begin{equation}
D_{n+1}\left( t\right)=d_{n+1}\left( t\right)\frac{\alpha^{n}}{\sqrt {n!}},
\end{equation}
\begin{equation}
B_{n+2}\left( t\right)=b_{n+2}\left( t\right)\frac{\alpha^{n}}{\sqrt {n!}}.
\end{equation}
The exact solution for the coupled differential equations is given by
\begin{equation}
a_{n}\left( t\right)= \sum\limits_{l = 1}^4 {\tilde{a}_{l}\exp\left( iM_{l}t\right)},
\end{equation}
\begin{equation}
c_{n+1}\left( t\right)= \sum\limits_{l = 1}^4 {\tilde{c}_{l}\exp\left( iM_{l}t\right)},
\end{equation}
\begin{equation}
d_{n+1}\left( t\right)= \sum\limits_{l = 1}^4 {\tilde{d}_{l}\exp\left( iM_{l}t\right)},
\end{equation}
\begin{equation}
b_{n+2}\left( t\right)= \sum\limits_{l = 1}^4 {\tilde{b}_{l}\exp\left( iM_{l}t\right)},
\end{equation}
where
\begin{widetext}
\begin{equation}
\tilde{a}_{l}=\frac{R_{1}+\xi_{1}\left(M_{k}+M_{m}+M_{n} \right)+R_{2}\left(M_{k}M_{m}+M_{k}M_{n}+M_{m}M_{n}\right)+ M_{k}M_{m}M_{n} }{M_{lk}M_{lm}M_{ln}},
\end{equation}
\begin{equation}
\tilde{c}_{l}=\frac{R_{3}+f_{1}\left(M_{k}+M_{m}+M_{n} \right)+R_{4}\left(M_{k}M_{m}+M_{k}M_{n}+M_{m}M_{n}\right)}{M_{lk}M_{lm}M_{ln}},
\end{equation}
\begin{equation}
\tilde{d}_{l}=\frac{R_{5}+f_{1}\left(M_{k}+M_{m}+M_{n} \right)+R_{6}\left(M_{k}M_{m}+M_{k}M_{n}+M_{m}M_{n}\right)}{M_{lk}M_{lm}M_{ln}},
\end{equation}
\begin{equation}
\tilde{b}_{l}=\frac{R_{7}+R_{8}\left(M_{k}M_{m}+M_{k}M_{n}+M_{m}M_{n}\right)}{M_{lk}M_{lm}M_{ln}}, \hspace{1cm} M_{lk}=M_{l}-M_{k}, 
\end{equation}
\end{widetext}
with
\begin{equation}
R_{1}=\xi_{1}^{3}+\left( f_{1}^{2}+f_{2}^{2}\right) \left(2\xi_{1}+\xi_{2}\right),
\end{equation}
where 
\begin{equation}
	{f_1} = {\lambda _2}\sqrt {n + 1}, \hspace{0.5cm} {f_2} = {\lambda _1}\sqrt {n + 1},
\end{equation}
and
\begin{equation}
R_{2}=f_{1}^{2}+f_{2}^{2}+\xi_{1}^{2},
\end{equation}
\begin{equation}
R_{3}=f_{2}\left(3 f_{1}^{2}+f_{2}^{2}+\xi_{1}^{2}+\xi_{1}\xi_{2}+\xi_{2}^{2}\right),
\end{equation}
\begin{equation}
R_{4}=2f_{1}f_{2}\left(\xi_{1}+\xi_{2}+\xi_{3}\right),
\end{equation}\begin{equation}
R_{5}=f_{1}\left( f_{1}^{2}+3f_{2}^{2}+\xi_{1}^{2}+\xi_{1}\xi_{2}+\xi_{2}^{2}\right),
\end{equation}
\begin{equation}
R_{6}=f_{1}\left(\xi_{1}+\xi_{2}\right),
\end{equation}
\begin{equation}
R_{7}=f_{1}\xi_{1}+f_{2}\xi_{2},
\end{equation}
\begin{equation}
R_{8}=2f_{1}f_{2}.
\end{equation}
The quantities $M_{l}$ ($l=1,2,3,4$) in the above equations are the roots of the following fourth-order equation
\begin{equation}
M^{4}+\chi_{3}M^{3}+\chi_{2}M^{2}+\chi_{1}M+\chi_{0}=0, \label{f_o_e}
\end{equation}
with
\begin{equation}
\chi_{3}=-\xi_{1}-2\xi_{2}-\xi_{3},
\end{equation}
\begin{equation}
\chi_{2}=2\xi_{2}\xi_{3}+\xi_{1}\xi_{3}+2\xi_{1}\xi_{2}+\xi_{2}^{2}-2f_{1}^{2}-2f_{2}^{2},
\end{equation}
\begin{align}
\chi_{1}=&f_{1}^{2}\xi_{1}+f_{2}^{2}\xi_{3}+2f_{1}^{2}\xi_{2}+2f_{2}^{2}\xi_{2}-\xi_{1}\xi_{2}^{2}\notag\\& +f_{1}^{2}\xi_{3}+f_{2}^{2}\xi_{3}-2\xi_{1}\xi_{2}\xi_{3}-\xi_{2}^{2}\xi_{3},
\end{align}
\begin{align}
\chi_{0}=&f_{1}^{4}-2f_{1}^{2}f_{2}^{2}+f_{2}^{4}-f_{1}^{4}\xi_{1}\xi_{2}-f_{2}^{2}\xi_{1}\xi_{2}\notag\\& -f_{1}^{2}\xi_{2}\xi_{3}-f_{2}^{2}\xi_{2}\xi_{3}+\xi_{1}\xi_{2}^{2}\xi_{3}.
\end{align}
The solutions of the equation (\ref{f_o_e}) write as
\begin{equation}
M_{1,2}=-\frac{\chi_{3}}{4}-\frac{g_{1}}{2}\pm\frac{Z_{-}}{2},
\end{equation}
and
\begin{equation}
M_{3,4}=-\frac{\chi_{3}}{4}+\frac{g_{1}}{2}\pm\frac{Z_{+}}{2},
\end{equation}
where
\begin{equation}
g_{1}=\sqrt{\frac{U_{1}}{3\upsilon_{1}}+\frac{\upsilon_{2}}{3}+U_{2}}, \hspace{1cm}Z_{\pm}=\sqrt{\eta_{2}\pm\frac{\eta_{3}}{4\eta_{1}}},
\end{equation}
\begin{equation}
\upsilon_{1}=27\chi_{1}^{2}-72\chi_{0}\chi_{2}+2\chi_{2}^{3}-g\chi_{1}\chi_{2}\chi_{3}+27\chi_{0}\chi_{3}^{2},
\end{equation}
\begin{equation}
\upsilon_{2}=\left(\frac{\upsilon_{1}+\sqrt{\upsilon_{1}^{2}-4U_{1}^{3}}}{2}\right)^{\frac{1}{3}},\hspace{0.5cm}\eta_{1}=\sqrt{U_{1}+\frac{U_{1}}{3\upsilon_{2}}+\frac{\upsilon_{2}}{3}},
\end{equation}
\begin{equation}
\eta_{2}=2U_{2}-\frac{U_{1}}{3\upsilon_{1}}-\frac{\upsilon_{2}}{3}, \hspace{0.7cm}\eta_{3}=-8\chi_{1}+4\chi_{2}\chi_{3}-\chi_{3}^{3}.
\end{equation}
In a special case,  we consider that ${c_{n + 1}}\left( t \right) = {d_{n + 1}}\left( t \right) = {e_{n + 1}}\left( t \right)$. In this case, the system of coupled differential equations reduced to
\begin{equation}
i\left[ {\begin{array}{*{20}{c}}
	{{{\mathop a\limits^\bullet}_n}\left( t \right)}\\
	{{{\mathop e\limits^\bullet}_{n + 1}}\left( t \right)}\\
	{{{\mathop b\limits^\bullet}_{n + 2}}\left( t \right)}
	\end{array}} \right] = \left[ {\begin{array}{*{20}{c}}
	{{\xi _1}}&{{f_1} + {f_2}}&0\\
	{{f_1}}&{{\xi _2}}&{{f_2}}\\
	0&{{f_1} + {f_2}}&{{\xi _4}}
	\end{array}} \right]\left[ {\begin{array}{*{20}{c}}
	{{a_n}\left( t \right)}\\
	{{e_{n + 1}}\left( t \right)}\\
	{{b_{n + 2}}\left( t \right)}
	\end{array}} \right].
\end{equation}
Solving the coupled differential equations obtained from the Schrödinger equation gives the following general solutions of the probability amplitudes
\begin{equation}
{b_{n + 2}}\left( t \right) = \sum\limits_{j = 1}^3 {{k_j}{\exp\left[ i{\chi _j}t\right] }},
\end{equation}
\begin{equation}
{e_{n + 1}}\left( t \right) =  - \frac{1}{{\left( {{f_1} + {f_2}} \right)}}\sum\limits_{j = 1}^3 {{k_j}\left( {{\chi _j} + {\xi _4}} \right){\exp\left[ i{\chi _j}t\right]}},
\end{equation}
and
\begin{eqnarray}
{a_n}\left( t \right) &=& \frac{1}{{{f_1}\left( {{f_1} + {f_2}} \right)}}\sum\limits_{j = 1}^3 {k_j}\left[ \left( {{\chi _j} + {\xi _4}} \right)\left( {{\chi _j} + {\xi _2}} \right)\right.\nonumber\\
		&~&\left. - {f_2}\left( {{f_1} + {f_2}} \right) \right]{\exp\left[i{\chi _j}t\right]}
\end{eqnarray}
where ${k_j}$ ($i=1,2,3$) are time-independent functions and ${\chi _j}$ are the eigenvalues of the matrix
\begin{equation}
W\left( t \right) = \left[ {\begin{array}{*{20}{c}}
	{{\xi _1}}&{{f_1} + {f_2}}&0\\
	{{f_1}}&{{\xi _2}}&{{f_2}}\\
	0&{{f_1} + {f_2}}&{{\xi _4}}
	\end{array}} \right]. \label{matrixM}
\end{equation}
It is simple to verify that the eigenvalues of the matrix  $W\left( t \right)$ (\ref{matrixM}) are analytically obtained as
\begin{equation}
{\chi _1} = \frac{1}{2}\left[ {{\xi _1} + {\xi _2} - \sqrt {{{\left( {{\xi _1} - {\xi _2}} \right)}^2} + 4{f_1}\left( {{f_1} + {f_2}} \right)} } \right],
\end{equation}
\begin{equation}
{\chi _2} = \frac{1}{2}\left[ {{\xi _1} + {\xi _2} + \sqrt {{{\left( {{\xi _1} - {\xi _2}} \right)}^2} + 4{f_1}\left( {{f_1} + {f_2}} \right)} } \right],
\end{equation}
and
\begin{equation}
{\chi _3} = {\xi _4}.
\end{equation}
From Eqs.(\ref{An})-(\ref{Dn}), we obtain the following conditions
\begin{equation}
{\begin{array}{*{20}{l}}
	{{k_1}{\chi _1}^2 + {k_2}{\chi _2}^2 + {k_3}{\chi _3}^2 = {f_1}\left( {{f_1} + {f_2}} \right),}\\
	{{k_1} + {k_2} + {k_3} = 0,}\\
	{{k_1}{\chi _1} + {k_2}{\chi _2} + {k_3}{\chi _3} = 0.}
	\end{array}}
\end{equation}

	


	
In this picture, the eigenvalues of the matrix $W\left( t \right)$ (\ref{matrixM}) as a function of the transition frequency ${w_z}$, the frequency of the mode of the cavity field ${w_c}$ and the photon number $n$ are
\begin{equation}
{\chi _1} = \frac{1}{2}\left[ {{w_z} + {w_c}\left( {2n + 1} \right) + \left( {{\gamma _1} + {\gamma _2}} \right)\left( {n + 2} \right) - \delta } \right],
\end{equation}
\begin{equation}
{\chi _2} = \frac{1}{2}\left[ {{w_z} + {w_c}\left( {2n + 1} \right) + \left( {{\gamma _1} + {\gamma _2}} \right)\left( {n + 2} \right) + \delta } \right],
\end{equation}
and
\begin{equation}
{\chi _3} =  - {w_z} + {w_c}\left( {n + 2} \right) - \left( {{\gamma _1} + {\gamma _2}} \right)\left( {n + 4} \right),
\end{equation}
where the function $\delta$ is defined by
\begin{equation}
\delta  = {\left( {{w_z} - {w_c} + \left( {{\gamma _1} + {\gamma _2}} \right)\left( {n + 2} \right)} \right)^2} + 4{\lambda _2}\left( {{\lambda _1} + {\lambda _2}} \right)\left( {n + 1} \right).
\end{equation}
After some straightforward calculations, we can obtain the functions ${k_1}$, ${k_2}$ and ${k_3}$ as
\begin{equation}
{k_1} = \frac{{2{\lambda _2}\left( {{\lambda _1} + {\lambda _2}} \right)\left( {n + 1} \right)}}{{\delta  - F\sqrt \delta  }},
\end{equation}
\begin{equation}
{k_2} = \frac{{2{\lambda _2}\left( {{\lambda _1} + {\lambda _2}} \right)\left( {n + 1} \right)}}{{\delta  + F\sqrt \delta  }},
\end{equation}
and
\begin{equation}
{k_3} = \frac{{4{\lambda _2}\left( {{\lambda _1} + {\lambda _2}} \right)\left( {n + 1} \right)}}{{{F^2} - \delta }},
\end{equation}
where the function $F$ is defined by
\begin{equation}
F = 3\left( {{w_z} - {w_c}} \right) + \left( {{\gamma _1} + {\gamma _2}} \right)\left( {3n + 10} \right).
\end{equation}
Using these results, the atom-atom density operator $\rho_{AB}$ can be obtained by tracing out the electromagnetic cavity field degrees of freedom. It is given by $\rho_{AB}=\sum\limits_{n'= 0}^\infty{\left\langle n'|{\psi}\left( t \right) \right\rangle \left\langle {\psi}\left( t \right)|n'\right\rangle}$. In the computational basis $\left\{ {\left| {{e_1},{e_2}} \right\rangle ,\left| {{e_1},{g_2}} \right\rangle ,\left| {{g_1},{e_2}} \right\rangle ,\left| {{g_1},{g_2}} \right\rangle } \right\}$, the density matrix $\rho_{AB}$ takes the form
\begin{equation}
\rho_{AB}  = \left( {\begin{array}{*{20}{c}}
	{{{\left| {{a_n}\left( t \right)} \right|}^2}}&0&0&0\\
	0&{{{\left| {{e_{n + 1}}\left( t \right)} \right|}^2}}&{{{\left| {{e_{n + 1}}\left( t \right)} \right|}^2}}&0\\
	0&{{{\left| {{e_{n + 1}}\left( t \right)} \right|}^2}}&{{{\left| {{e_{n + 1}}\left( t \right)} \right|}^2}}&0\\
	0&0&0&{{{\left| {{b_{n + 2}}\left( t \right)} \right|}^2}}
	\end{array}} \right),\label{Matrxgho}
\end{equation}
which is a two-qubit state of $X$ type and their entries are given by
\begin{widetext}
\begin{align}
{\left| {{e_{n + 1}}\left( t \right)} \right|^2} =& {\lambda _2}^2\left( {n + 1} \right)\left[ {\sum\limits_ \pm  {\frac{{ - {w_z} + {w_c}\left( {4n + 5} \right) - \left( {{\gamma _1} + {\gamma _2}} \right)\left( {n + 6} \right) \mp \sqrt \delta  }}{{{{\left( {\delta  \mp F\sqrt \delta  } \right)}^2}}}}  + \frac{{64{{\left( { - {w_z} + {w_c}\left( {n + 2} \right) - \left( {{\gamma _1} + {\gamma _2}} \right)\left( {n + 4} \right)} \right)}^2}}}{{{{\left( {{F^2} - \delta } \right)}^2}}}} \right] \notag\\&+ \frac{{2{\lambda _2}^2\left( {n + 1} \right)\cos \left( {\sqrt \delta  t} \right)}}{{{\delta ^2} - F\delta }}\left[ {{{\left( { - {w_z} + {w_c}\left( {4n + 5} \right) - \left( {{\gamma _1} + {\gamma _2}} \right)\left( {n + 6} \right)} \right)}^2} - \delta } \right]+ \notag\\& + \sum\limits_ \pm  {\frac{{16{\lambda _2}^2\left( {n + 1} \right)\cos \left( {\left( {\frac{{F \mp \sqrt \delta  }}{2}} \right)t} \right)}}{{\left( {{F^2} - \delta } \right)\left( {\delta  \mp F\sqrt \delta  } \right)}}\left[ \begin{array}{l}
	\left( { - {w_z} + {w_c}\left( {n + 2} \right) - \left( {{\gamma _1} + {\gamma _2}} \right)\left( {n + 4} \right)} \right) \times \\
	\left( { - {w_z} + {w_c}\left( {4n + 5} \right) - \left( {{\gamma _1} + {\gamma _2}} \right)\left( {n + 6} \right) \mp \sqrt \delta  } \right)
	\end{array} \right]},
	\end{align}
\begin{align}
{\left| {{b_{n + 2}}\left( t \right)} \right|^2} = \frac{{8{\lambda _2}^2{{\left( {{\lambda _1} + {\lambda _2}} \right)}^2}{{\left( {n + 1} \right)}^2}}}{{\left( {\delta  - {F^2}} \right)}}\left[ {\frac{{3\delta  + {F^2}}}{{\delta \left( {\delta  - {F^2}} \right)}} + \frac{{\cos \left( {\sqrt \delta  t} \right)}}{\delta } - \sum\limits_ \pm  {\frac{{\cos \left( {\frac{{F \mp \sqrt \delta  }}{2}t} \right)}}{{\delta  \mp F\sqrt \delta  }}} } \right],
\end{align}
and	
\begin{equation}
{\left| {{a_n}\left( t \right)} \right|^2} = 1 - 2{\left| {{e_{n + 1}}\left( t \right)} \right|^2} - {\left| {{b_{n + 2}}\left( t \right)} \right|^2}.
\end{equation}
\end{widetext}
Having determined the bipartite atomic density matrix operator, we shall investigate the quantum coherence and quantum discord. In particular, we shall investigate the influence of effect of the stark-shift parameter on these quantum measures.
\section{Quantum Coherence} \label{sec:QC}
Arising from the superposition principle, quantum coherence (QC) is a key concept to understand the weirdness of the fundamental aspects of quantum mechanics. QC is an important resource in different quantum information processing tasks. The characterization of QC in composite quantum systems and in particular how it can be quantified is one of the challenges facing both quantum information theory and quantum technologies. For this reason, a rigorous framework to quantify the degree of superposition in quantum states has been established by Åberg \cite{Aberg2006,Aberg2014} and recently updated by Baumgratz et al \cite{Baumgratz2014}. By analogy with the resource theory of quantum entanglement which establishes the sets of separable states and local operations and classical communication (LOCC), the quantification of QC is based on the concepts of incoherent states and incoherent operations. For a quantum state associated with a $d$-dimensional Hilbert space ${\cal H}$, we fix the orthogonal basis $\left\{ {\left| j \right\rangle } \right\}_{j = 1}^d$ as the reference basis. If a quantum state is diagonal when expressed on this local reference basis, it is called incoherent state. On the other hand, an incoherent operation is a completely positive and trace-preserving linear map (CPTP) that maps an incoherent state to an incoherent state and no creation of coherence could be observed.
\par
The Baumgratz et al analysis \cite{Baumgratz2014} has attracted deep attention of many researchers and various measures of QC, which satisfy the physical requirements of noncontractivity and monotonicity, have been proposed since then. We quote, for instance, the $l_{1}$-norm coherence \cite{Baumgratz2014}, the coherence of formation \cite{Yu2016,Winter2016}, the relative entropy coherence \cite{Baumgratz2014}, the geometric measure of coherence \cite{Streltsov2015}, the distillable coherence \cite{Winter2016,Yuan2015}, the coherence measures based on entanglement \cite{Streltsov2015,Qi2017}, the coherence measures based on trace norm \cite{Yu2016}, the coherence measure via quantum skew information \cite{Yu2017}, via Tsallis relative entropy \cite{Rastegin2016,Zhao2018}, via fidelity \cite{Liu2017} and via relative entropy in Gaussian states \cite{Xu2016}.
\par
It is worth noticing that all of the QC measures of a quantum state can be classified into two categories; the first concerns the entropic measures which are based on the entropic function and the second situation concerns the geometric class of the measures which has a metric character and is quantified as its distance to the closest incoherent state. Recently, a new measure of CQ  based on the quantum version of the Jensen-Shannon divergence has been introduced by Radhakrishnan et al \cite{Radhakrishnan2016}. This measure has both entropic and geometric characters and is easy to compute analytically for a generic quantum state. In addition, it satisfies the full physical requirements that every good QC quantifier should satisfy \cite{Baumgratz2014}. The quantum Jensen-Shannon divergence, as a measure of the distance between two quantum states,  is defined as \cite{Lin1991,Briet2009,Majtey2005,Lamberti2008}
\begin{equation}
{\mathfrak{J}}\left( {\rho ,{\sigma}} \right) = S\left( {\frac{{\rho  + {\sigma}}}{2}} \right) - \frac{1}{2}\left( {S\left( \rho  \right) + S\left( {{\sigma}} \right)} \right),
\end{equation}
and the QC is defined as the square root of the quantum Jensen-Shannon divergence, i.e.,
\begin{equation}
{\cal C}\left( \rho  \right) = \sqrt {\mathfrak{J}\left( {\rho ,{\rho _d}} \right)}, \label{EQC}
\end{equation}
where $S\left(\rho\right)=-{\rm Tr}\left( {\rho \log_{2}\rho } \right)$ is the von Neumann entropy and $\rho _d$ is the diagonal part of quantum state $\rho$ in the computational basis.\par
Hence, the expression of QC for two 2-level atoms interacting with a single-mode electromagnetic cavity field in the presence of the Stark-shift effect can be easily computed. Indeed, using the density matrix (\ref{Matrxgho}), one can check that the quantum coherence defined by (\ref{EQC}) is given by
\begin{equation}
{\cal C}\left( \rho \right) = \sqrt {3{{\left| {{e_{n + 1}}} \right|}^2}\left( {1 - \frac{{{{\log }_2}\left( 3 \right)}}{2}} \right)}.
\end{equation}
\begin{widetext}
	
\begin{figure}[h]
	{{\begin{minipage}[b]{.30\linewidth}
				\centering
				\includegraphics[scale=0.42]{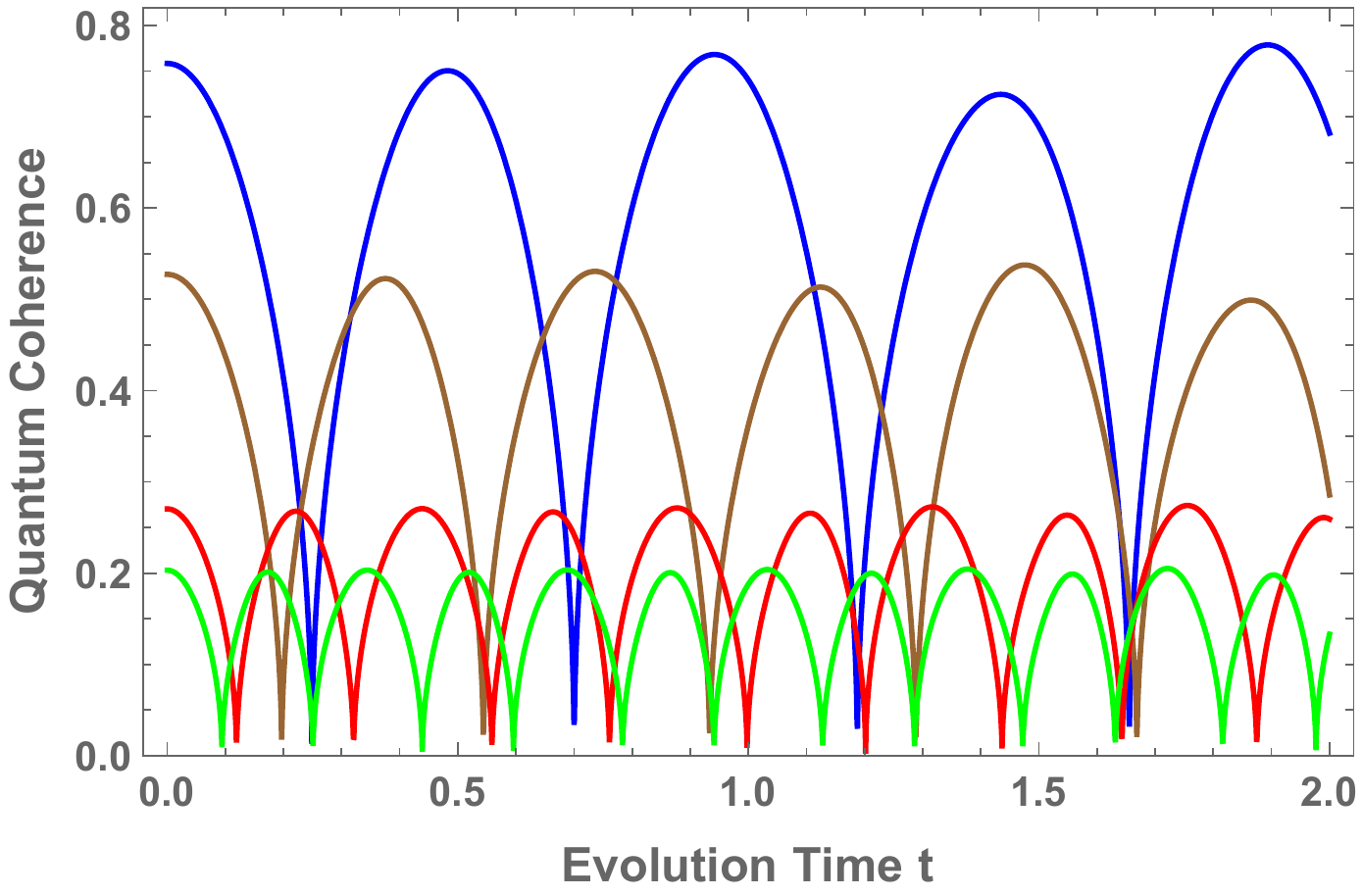} \vfill  $\left(A \right)$
			\end{minipage}\hfill
			\begin{minipage}[b]{.30\linewidth}
				\centering
				\includegraphics[scale=0.42]{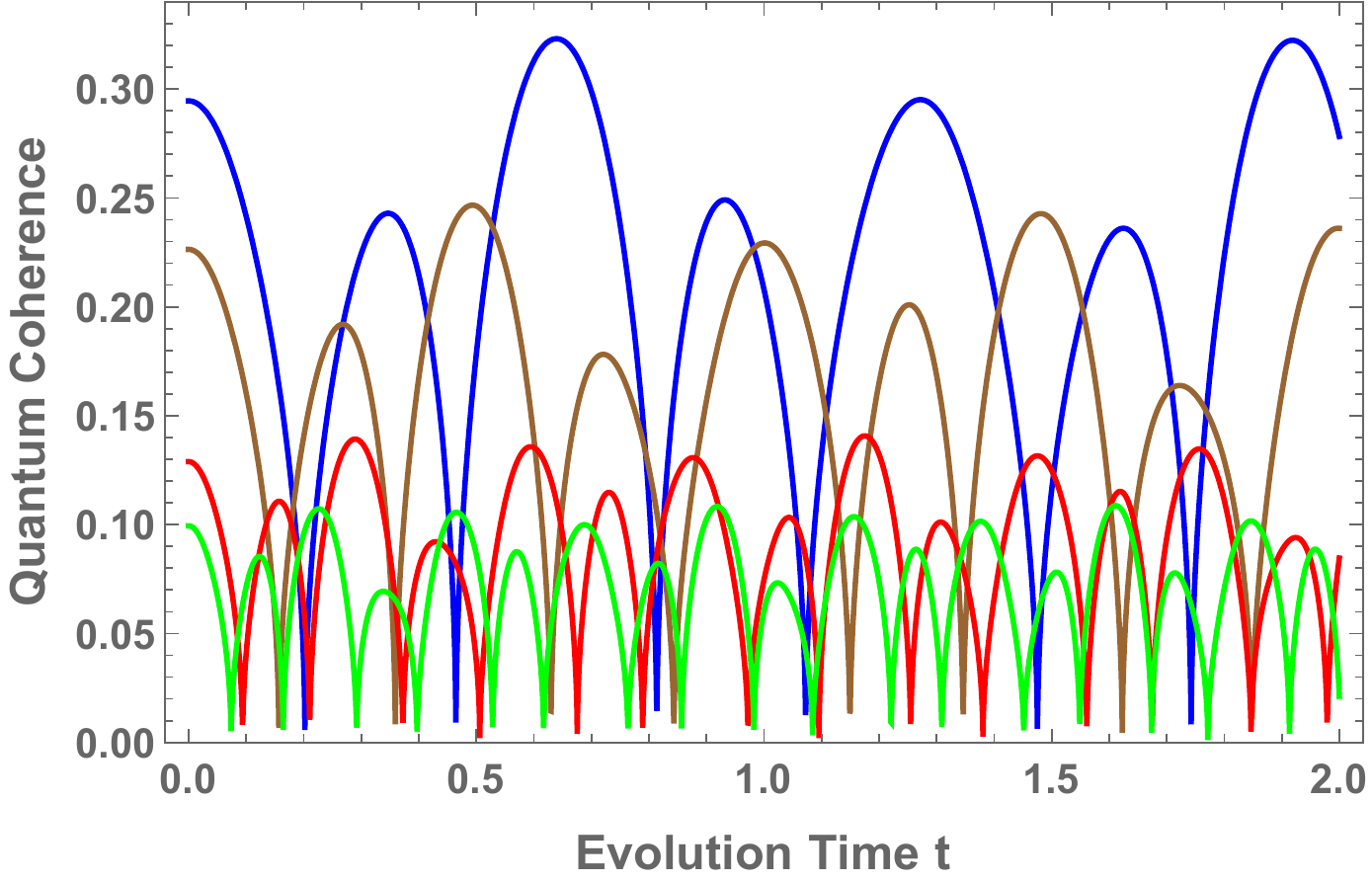} \vfill  $\left(B \right)$
			\end{minipage}\hfill
			\begin{minipage}[b]{.30\linewidth}
				\centering
				\includegraphics[scale=0.42]{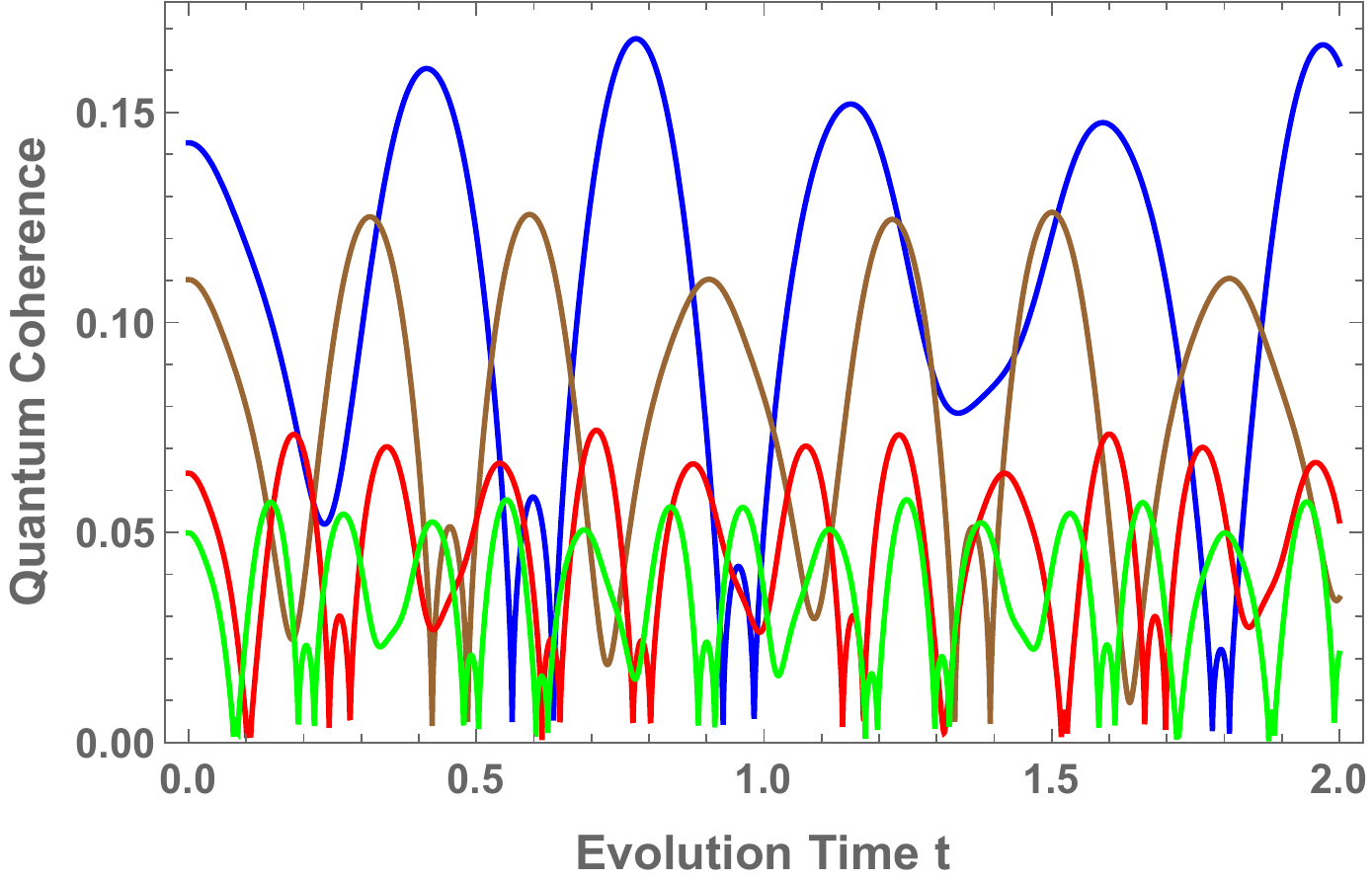} \vfill  $\left(C \right)$
	\end{minipage}}}
\caption{Time evolution of quantum coherence in terms of the time $t$ for different values of the Stark shift; $\gamma_1=1$ and $\gamma_2=2$ for blue curve, $\gamma_1=2$ and $\gamma_2=3$ for brown curve, $\gamma_1=3$ and $\gamma_2=4$ for red curve, $\gamma_1=4$ and $\gamma_2=5$ for green curve and for different values of the number of coherent state photons; $n=0$ for panel (A), $n=1$ for panel (B), $n=3$ for panel (C), when $w_{z}=w_{c}=0$.}\label{fig1}
\end{figure}
\begin{figure}[h]
	{{\begin{minipage}[b]{.30\linewidth}
				\centering
				\includegraphics[scale=0.42]{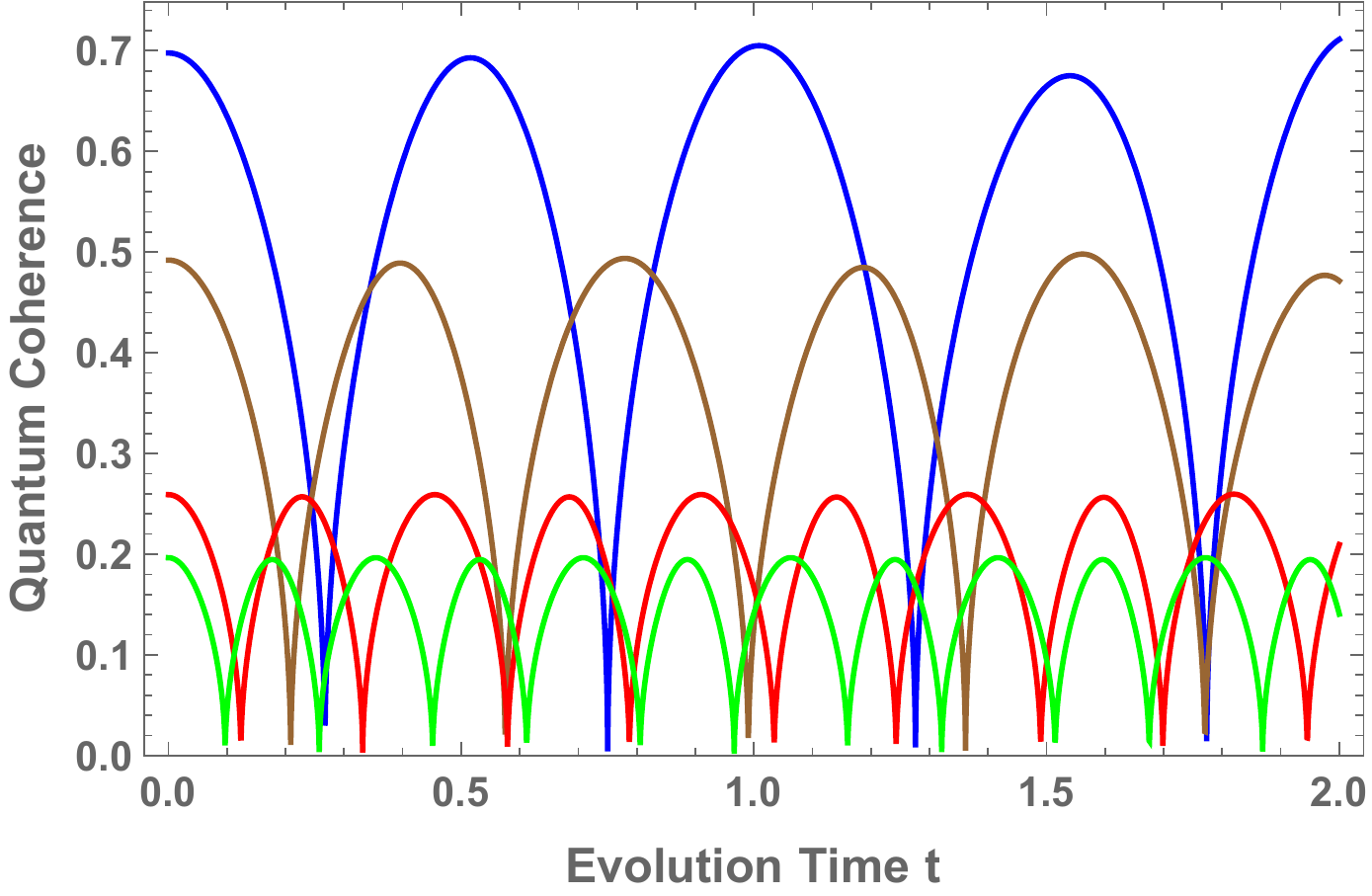} \vfill $\left(A\right)$
			\end{minipage}\hfill
			\begin{minipage}[b]{.30\linewidth}
				\centering
				\includegraphics[scale=0.42]{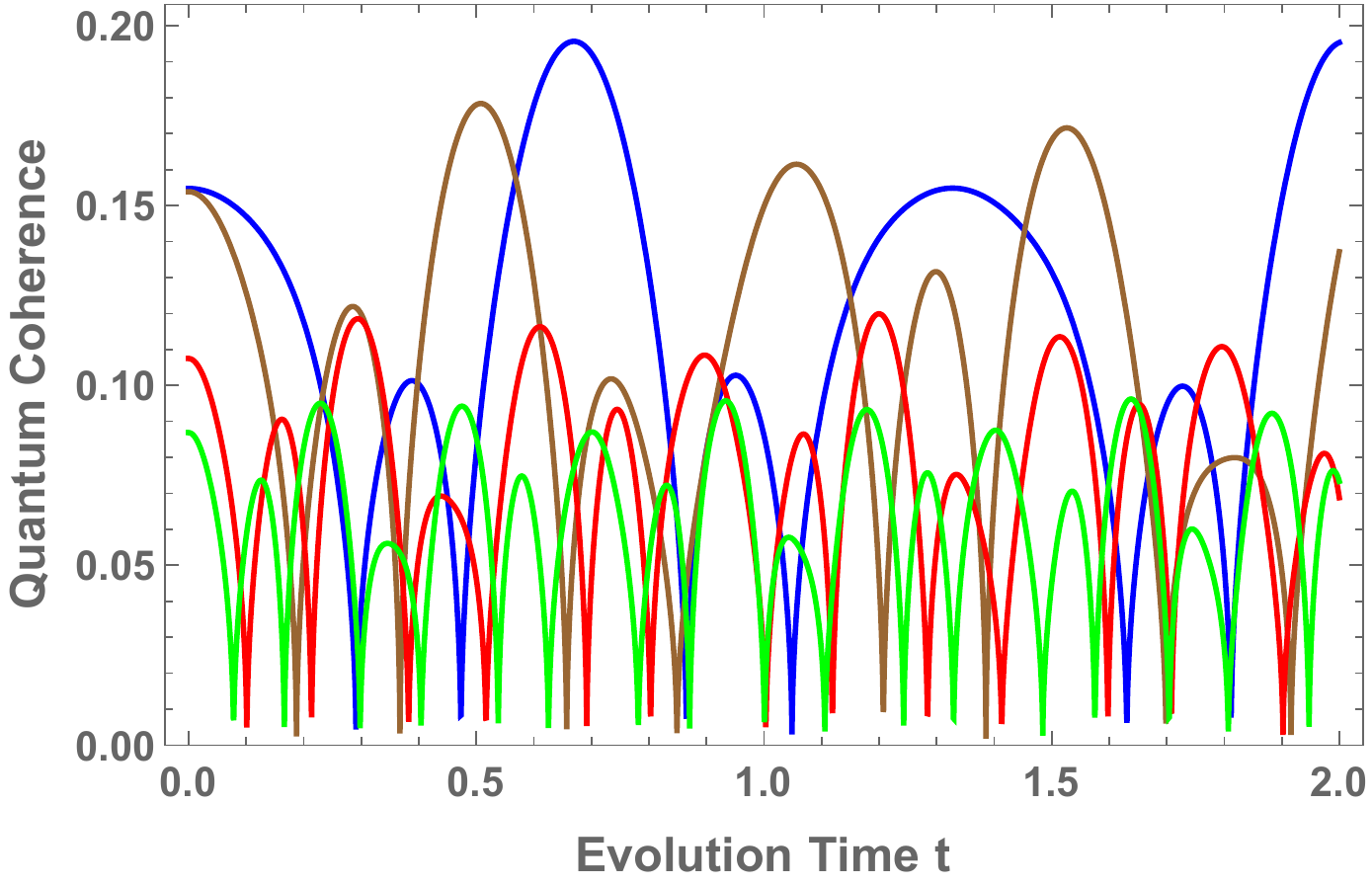} \vfill  $\left(B\right)$
			\end{minipage}\hfill
			\begin{minipage}[b]{.30\linewidth}
				\centering
				\includegraphics[scale=0.42]{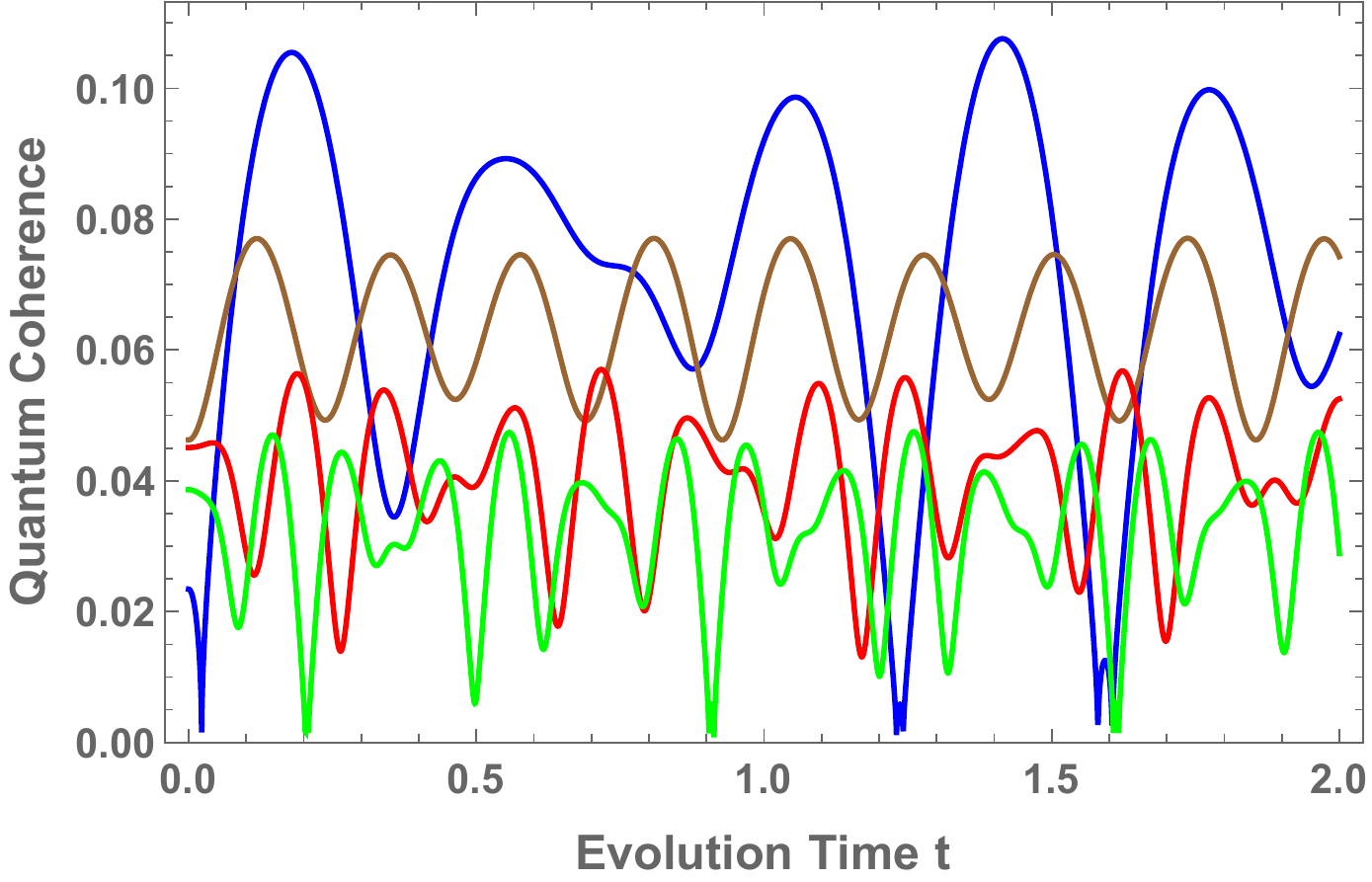} \vfill $\left(C \right)$
	\end{minipage}}}
	\caption{The same as in figure 2 but for different parameters of $w_{z}=0.5$ and $w_{c}=1$.}\label{fig2}
\end{figure}	
\end{widetext}
In order to analyze in detail the influence of the Stark shift on the QC dynamics of two two-level atoms interacting with a single-mode electromagnetic cavity field, we display the evolution of QC based on the quantum Jensen-Shannon divergence versus the time $t$ for various values of the Stark shift parameters in Fig.(\ref{fig1}). To simplify the numerical analysis, we consider the situation where $w_{z}$ (transition frequency) and $w_{c}$ (mode field cavity frequency) vanishes. From the plot of Fig.(\ref{fig1}), it is observed that the QC behavior of this model is a periodic function of time $t$ and their periods decrease with the increase of the Stark shift parameters ${\gamma_{1,2}}$. Also, the increase of the Stark shift parameters in the cavity leads to a rapid decrease of QC due to the intrinsic decoherence and the two atoms can remain in a stationary incoherent state for large values of the Stark shift parameters. As depicted in Fig.(\ref{fig1}), the amount of QC decreases over a certain interval of time showing the degradation of the superposition of the initial state. After, it increases and survives which shows the revival phenomenon of QC in the system. This revival phenomenon can be explained by the transfer of QC between the total system including the cavity field and the two atoms. Furthermore, we noticed also that the decrease in the amount of QC becomes more pronounced when passing from $n=0$ (Fig.(\ref{fig1}-A)) to $n=3$ (Fig.(\ref{fig1}-C)). This means that QC decreases abruptly with the increasing number of photons. In Fig.(\ref{fig2}), we depict the behavior of QC versus the time $t$ for different values of the Stark shift parameters with $w_{z}=0.5$ and $w_{c}=1$. These results are in agreement with the results obtained in Fig.(\ref{fig1}) which shows that for fixed Stark shift parameters, the QC decreases as the number of coherent state photons increases. Also, for a fixed number of coherent state photons, the QC decreases as the Stark shift parameters increases. On the other hand, the results reported in Fig.(\ref{fig2}) show that increasing both the transition frequency and frequency of the mode of the cavity field destroys the amount of QC in the system and modify their periodicity. 

\section{Quantum Discord}\label{sec:QD}
In order to investigate the dynamic evolutions of the quantum correlations in our system, one can use the quantum discord (QD) measure which characterizes the nonclassicality of correlations in multipartite quantum system \cite{Ollivier2001,Henderson2001}. It is noteworthy that QD is captured by strong measurements that lead to the loss of its coherence, but it reveals more quantum correlations than entanglement in the sense that it may not disappear even for mixed separable quantum states. It is defined as the difference between two classically-equivalent expressions of the mutual information; the original quantum mutual information and the local measurement-induced quantum mutual information. According to ref.\cite{Ollivier2001}, the QD $Q\left( {\rho_{AB}} \right)$ in a two-qubit state $\rho_{AB}$ is defined by:
\begin{equation}
{\cal Q}\left( {{\rho _{AB}}} \right): = {\cal I}\left( {{\rho _{AB}}} \right) -{\cal J}\left( {{\rho _{AB}}} \right), \label{QDD}
\end{equation}
with the total correlation is quantified by the quantum mutual information ${\cal I}\left( {{\rho_{AB}}} \right): = S\left( {{\rho_{A}}} \right) + S\left( {{\rho_{B}}} \right) - S\left( {{\rho_{AB}}} \right)$, and the quantity ${\cal J}\left( {{\rho _{AB}}} \right)$ is defined as a measure of classical correlation
\begin{equation}
{\cal J}\left( {{\rho _{AB}}} \right)=\mathop {\max }\limits_{{\pi _B}^j} \left( {S\left( {{\rho _B}} \right) - \sum\limits_j {{p_{B,j}}S\left( {{\rho _{^{B,j}}}} \right)} } \right), \label{QCD}
\end{equation}
where the minimum is taken over the set of positive operator valued measurements (POVM) $\{{\pi _B}^j\}$ on subsystem $B$ which satisfy $\sum\limits_j {{\pi _B}{{^j}^\dag }{\pi _B}^j = I} $, $S\left(\rho\right)$ is the von Neumann entropy, ${p_{B,j}} = {\rm Tr}_{AB}\left[\left( {I \otimes {\pi _B}^j} \right){{\rho _{AB}}\left({I \otimes {\pi _B}^{j^{\dagger}}} \right)}\right]$ and ${\rho _{B,j}} ={{{\rm Tr}_A\left[ {\left( {I \otimes {\pi _B}^j} \right){\rho _{AB}}\left( {I \otimes {\pi _B}^{j^{\dagger}}}\right)} \right]} \mathord{\left/
		{\vphantom {{{\rm Tr}_A\left[ {\left( {I \otimes {\pi _B}^j} \right){\rho _{AB}}\left( {I \otimes {\pi _B}^{j^{\dagger}}}\right)} \right]} {p_{B,j}}}} \right.
		\kern-\nulldelimiterspace} {p_{B,j}}}$ are the probability and the conditional state of subsystem $B$ associated with outcome $j$. It is worth pointing out that the main idea of calculating the QD is to quantify the amount of information that is not accessible by a local measurement by extracting some information about subsystem $A$ by reading the state of subsystem $B$ without disturbing state $A$. The difficult step in evaluating QD is to find an analytical expression of classical correlations because it requires a minimization process in optimizing the conditional entropy over all possible local measurements. Due to this fact, the analytical derivation of QD may be performed only for some specific two-qubit \cite{Dakic2010,Girolami2011}. For the bipartite system under consideration in this paper, the analytic expressions of QD can be computed by using the method reported by C.Z Wang et al in \cite{Wang2010}. First, for a two qubit state of $X$-type, the density matrix has taken the following form
\begin{equation}
{\rho _{AB}} = \left( {\begin{array}{*{20}{c}}
	{{\rho _{11}}}&0&0&{{\rho _{14}}}\\
	0&{{\rho _{22}}}&{{\rho _{23}}}&0\\
	0&{{\rho _{23}}^ * }&{{\rho _{33}}}&0\\
	{{\rho _{14}}^ * }&0&0&{{\rho _{44}}}
	\end{array}} \right), \label{X-state}
\end{equation}
and the corresponding eigenvalues are given by
\begin{equation*}
{\eta _{1,2}} = \frac{1}{2}\left[ {{\rho _{11}} + {\rho _{44}} \pm \sqrt {{{\left( {{\rho _{11}} - {\rho _{44}}} \right)}^2} + 4{{\left| {{\rho _{14}}} \right|}^2}} } \right],
\end{equation*}
and
\begin{equation*}
{\eta _{3,4}} = \frac{1}{2}\left[ {{\rho _{22}} + {\rho _{33}} \pm \sqrt {{{\left( {{\rho _{22}} - {\rho _{33}}} \right)}^2} + 4{{\left| {{\rho _{23}}} \right|}^2}} } \right].
\end{equation*}
The entropy of the reduced matrix of $\rho_{A}$ and $\rho_{B}$ is given by
\begin{align}
S\left( {{\rho _A}} \right) =&  - \left( {{\rho _{11}} + {\rho _{22}}} \right){\log _2}\left( {{\rho _{11}} + {\rho _{22}}} \right) \notag\\&- \left( {{\rho _{33}} + {\rho _{44}}} \right){\log _2}\left( {{\rho _{33}} + {\rho _{44}}} \right),
\end{align}
and
\begin{align}
S\left( {{\rho _B}} \right) =&  - \left( {{\rho _{11}} + {\rho _{33}}} \right){\log _2}\left( {{\rho _{11}} + {\rho _{33}}} \right) \notag\\&- \left( {{\rho _{22}} + {\rho _{44}}} \right){\log _2}\left( {{\rho _{22}} + {\rho _{44}}} \right).
\end{align}
To minimize the classical correlations (\ref{QCD}), we take a complete set of projective measures on the subsystem $B$ which are represented by the operators $\left\{ {{\pi _B}^j = \left| {{B_j}} \right\rangle \left\langle {{B_j}} \right|,j = 1,2} \right\}$ with
\begin{equation}
\begin{array}{l}
\left| {{B_1}} \right\rangle  = \cos \left( {\frac{\theta }{2}} \right)\left| 1 \right\rangle  + {e^{i\varphi }}\sin \left( {\frac{\theta }{2}} \right)\left| 0 \right\rangle \\
\left| {{B_2}} \right\rangle  = \sin \left( {\frac{\theta }{2}} \right)\left| 1 \right\rangle  - {e^{i\varphi }}\cos \left( {\frac{\theta }{2}} \right)\left| 0 \right\rangle,
\end{array}
\end{equation}
with $0 \le \theta  \le \frac{\pi }{2}$ and $0 \le \varphi  \textless 2\pi$. The probability ${p_{B,j}}$ corresponding to the result $j$ and the two eigenvalues of the corresponding ${\rho _{^{B,j}}}$ after the measurement are given by
\begin{equation}
{p_{B,j}} = \frac{1}{2}\left[ {1 + {{\left( { - 1} \right)}^j}\cos \theta \left( {1 - 2{\rho _{11}} - 2{\rho _{33}}} \right)} \right]
\end{equation}
and
\begin{equation}
{\eta_ \pm }\left( {{\rho _{B,j}}} \right) = \frac{1}{2}\left( {1 \pm \frac{{\sqrt {{\upsilon_j}} }}{{{p_{B,j}}}}} \right),
\end{equation}
with
\begin{align}
{\upsilon _j} = &\frac{1}{4}{\left[ {1 - 2\left( {{\rho _{33}} + {\rho _{44}}} \right) + {{\left( { - 1} \right)}^j}\cos \theta \left( {1 - 2{\rho _{11}} - 2{\rho _{44}}} \right)} \right]^2} + \notag\\&{\sin ^2}\theta \left[ {{{\left| {{\rho _{14}}} \right|}^2} + {{\left| {{\rho _{23}}} \right|}^2} - 2\left| {{\rho _{14}}} \right|\left| {{\rho _{23}}} \right|\sin \left( {2\varphi  + \phi } \right)} \right].
\end{align}
The entropy of ${{\rho _{B,j}}}$ can be written in terms of its eigenvalues as follows
\begin{equation}
S\left( {{\rho _{B,j}}} \right) = H\left( {{\eta_+ }\left( {{\rho _{B,j}}} \right)} \right)
\end{equation}
where $H\left( x  \right) =  - x {\log _2}x  - \left( {1 - x } \right){\log _2}\left( {1 - x} \right)$ is the binary Shannon entropy. Thus, the conditional entropy is
\begin{equation}
S\left( {{\rho _{A\left| B \right.}}} \right) = \sum\limits_{j = 1}^2 {{p_{B,j}}S\left( {{\rho _{B,j}}} \right)}  = {p_{B,1}}S\left( {{\rho _{B,1}}} \right) + {p_{B,2}}S\left( {{\rho _{B,2}}} \right).
\end{equation}
By setting partial derivatives of this conditional entropy with respect to $\theta $ and $\varphi$ equal to zero, it is easy to show that the conditional entropy is minimal when $\theta  = \frac{\pi }{2}$ ( in this case ${p_{B,1}} = {p_{B,2}}$ and $S\left( {{\rho _{B,1}}} \right) = S\left( {{\rho _{B,2}}} \right)$). Thus, the minimum value of the conditional entropy is
\begin{equation}
{\zeta _1} = H\left( {\frac{1 + \sqrt {{{\left( {1 - 2\left( {{\rho _{33}} + {\rho _{44}}} \right)} \right)}^2} + 4{{\left( {\left| {{\rho _{14}}} \right| + \left| {{\rho _{23}}} \right|} \right)}^2}}}{2}} \right),
\end{equation}
The second extremal value is obtained when $\theta=0$ for any arbitrary value of $\varphi$ and the second minimal value of the conditional entropy is
\begin{equation}
{\zeta_2} = - \sum\limits_{i = 1}^4 {{\rho _{ii}}{{\log }_2}} {\rho _{ii}} - H\left( {{\rho _{11}} + {\rho _{33}}} \right).
\end{equation}
Consequently, the explicit expression of classical correlations for the $X$-states (\ref{X-state}) takes the form
\begin{equation}
{\cal J}\left( {{\rho _{AB}}} \right) = \max \left( {{{\cal J}_1},{{\cal J}_2}} \right),
\end{equation}
with
\begin{equation}
{{\cal J}_i} = H\left( {{\rho _{11}} + {\rho _{22}}} \right) - {\zeta _i}.
\end{equation}
Finally, the analytical expression of QD can be determined from Eq.(\ref{QDD}) as
\begin{equation}
{\cal Q}\left( {{\rho _{AB}}} \right) = \min \left( {{{\cal Q}_1},{{\cal Q}_2}} \right),\label{QDANLY}
\end{equation}
where
\begin{equation}
{{\cal Q}_i} = H\left( {{\rho _{11}} + {\rho _{33}}} \right) + \sum\limits_{i = 1}^4 {{\eta_i}{{\log }_2}} {\eta_i} + {\zeta _i}.
\end{equation}
For the two-qubit state (\ref{Matrxgho}), the quantum discord is given by the expression (\ref{QDANLY}) with
\begin{align}
{\cal Q}_{1} &= H\left( {1 - {{\left| {{e_{n + 1}}} \right|}^2} - {{\left| {{b_{n + 2}}} \right|}^2}} \right) + {\left| {{a_n}} \right|^2}{\log _2}\left( {{{\left| {{a_n}} \right|}^2}} \right) \notag\\&+ {\left| {{b_{n + 2}}} \right|^2}{\log _2}\left( {{{\left| {{b_{n + 2}}} \right|}^2}} \right) - 2{\left| {{e_{n + 1}}} \right|^2}{\log _2}\left( {2{{\left| {{e_{n + 1}}} \right|}^2}} \right) +\notag\\& H\left( {\frac{1}{2}\left( {1 + \sqrt {{{\left( {1 - 2{{\left| {{e_{n + 1}}} \right|}^2} - 2{{\left| {{b_{n + 2}}} \right|}^2}} \right)}^2} + 4{{\left| {{e_{n + 1}}} \right|}^4}} } \right)} \right),
\end{align}
and
\begin{equation}
{\cal Q}_{2}=2{{\left| {{e_{n + 1}}} \right|}^2}.
\end{equation}

\begin{widetext}
	
\begin{figure}[h]
	{{\begin{minipage}[b]{.30\linewidth}
				\centering
				\includegraphics[scale=0.42]{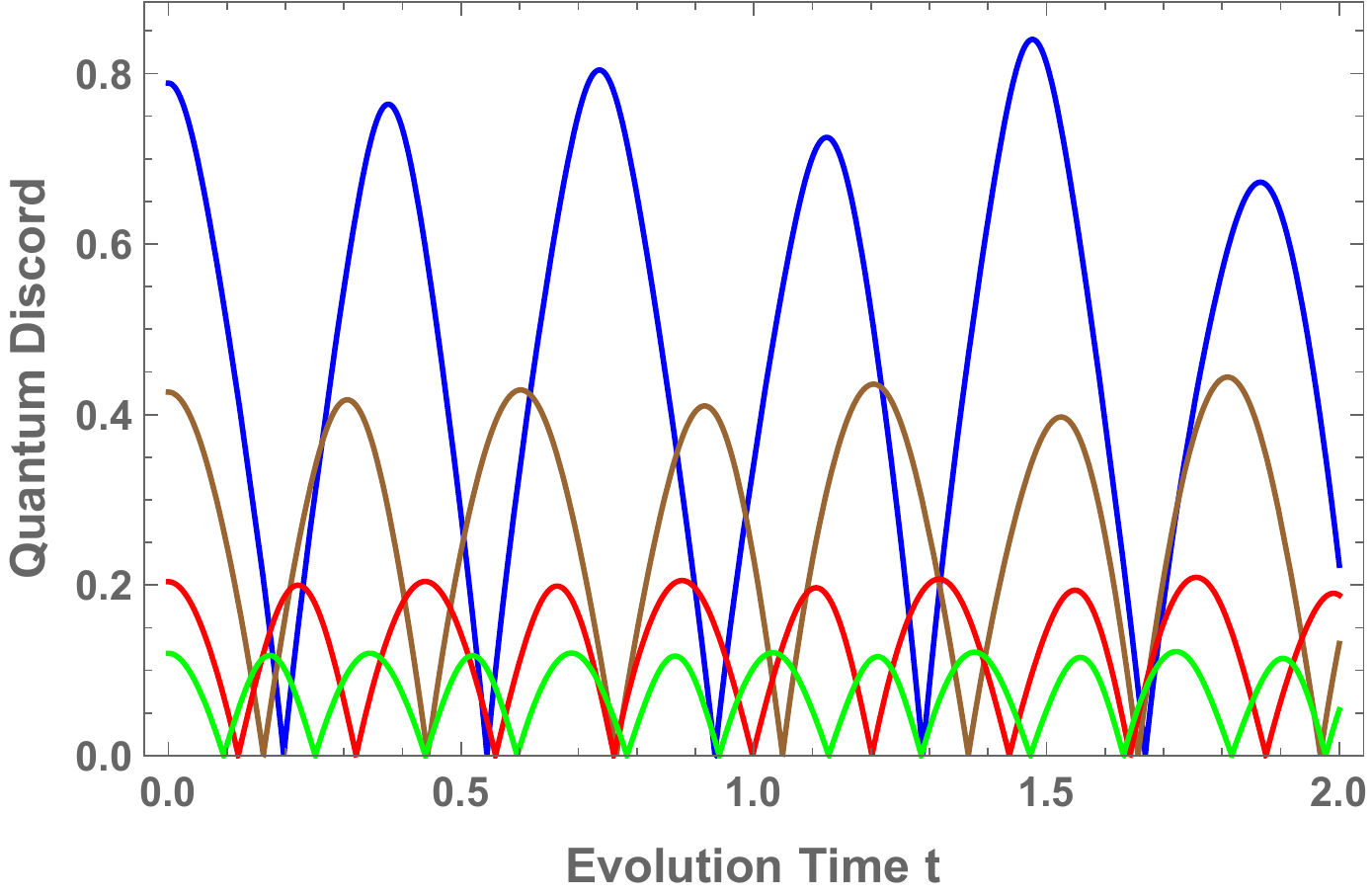} \vfill  $\left(A \right)$
			\end{minipage}\hfill
			\begin{minipage}[b]{.30\linewidth}
				\centering
				\includegraphics[scale=0.42]{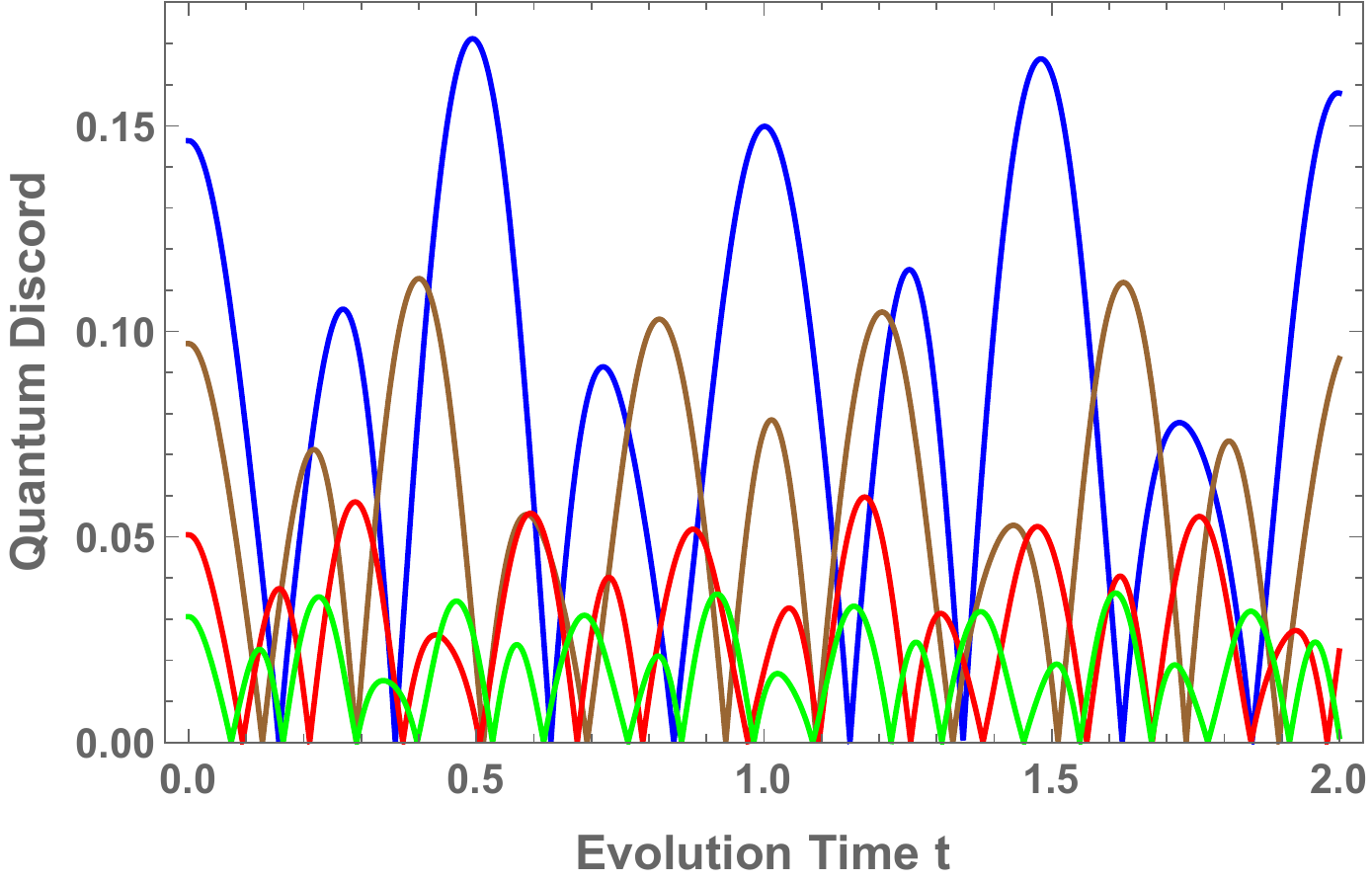} \vfill  $\left(B \right)$
			\end{minipage}\hfill
			\begin{minipage}[b]{.30\linewidth}
				\centering
				\includegraphics[scale=0.42]{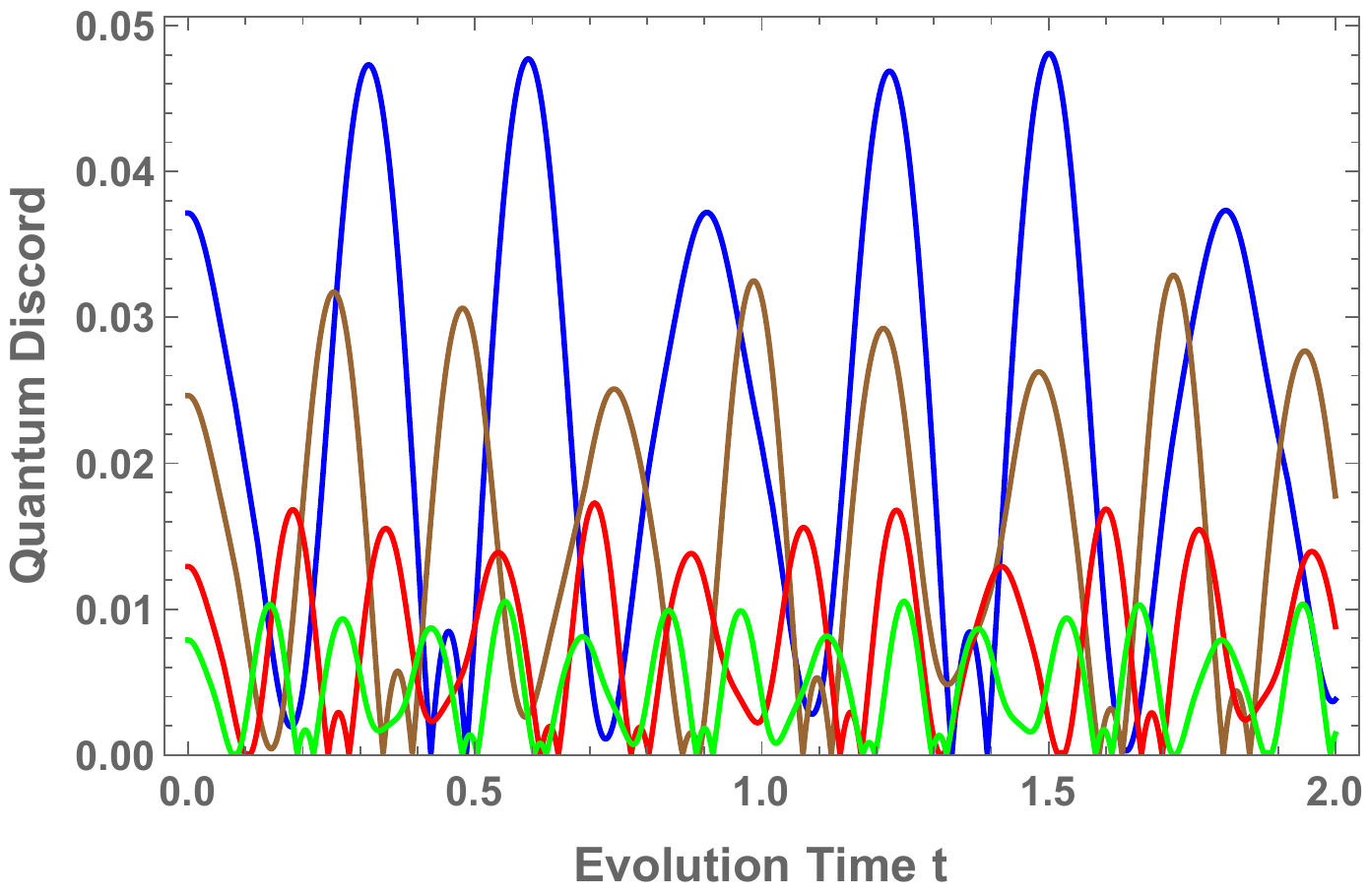} \vfill $\left(C \right)$
	\end{minipage}}}
\caption{The same as in Figure 2 but for the time evolution of quantum discord in terms of the time $t$ }\label{fig3}
\end{figure}
\begin{figure}[h]
	{{\begin{minipage}[b]{.30\linewidth}
				\centering
				\includegraphics[scale=0.42]{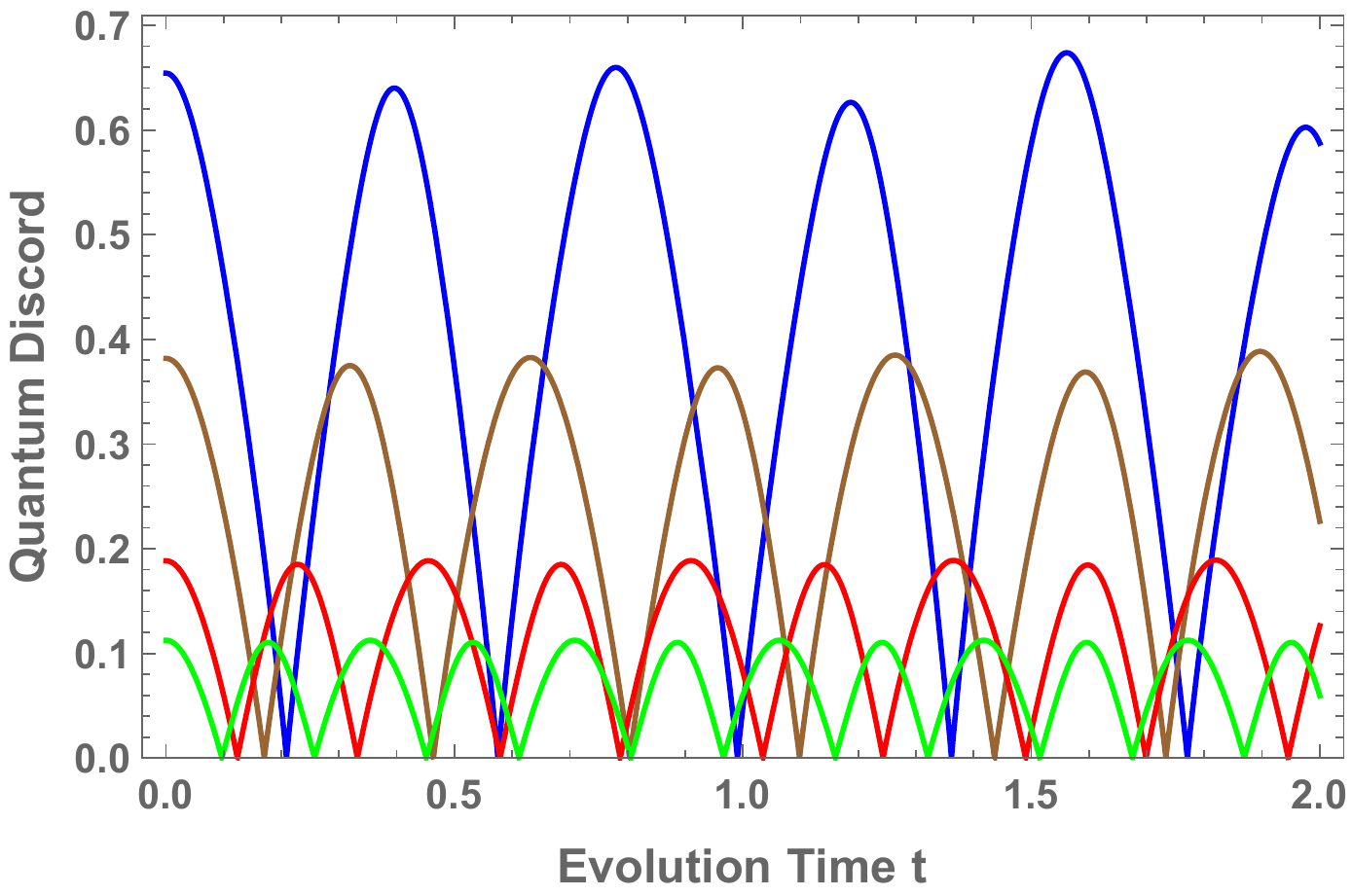} \vfill  $\left(A \right)$
			\end{minipage}\hfill
			\begin{minipage}[b]{.30\linewidth}
				\centering
				\includegraphics[scale=0.42]{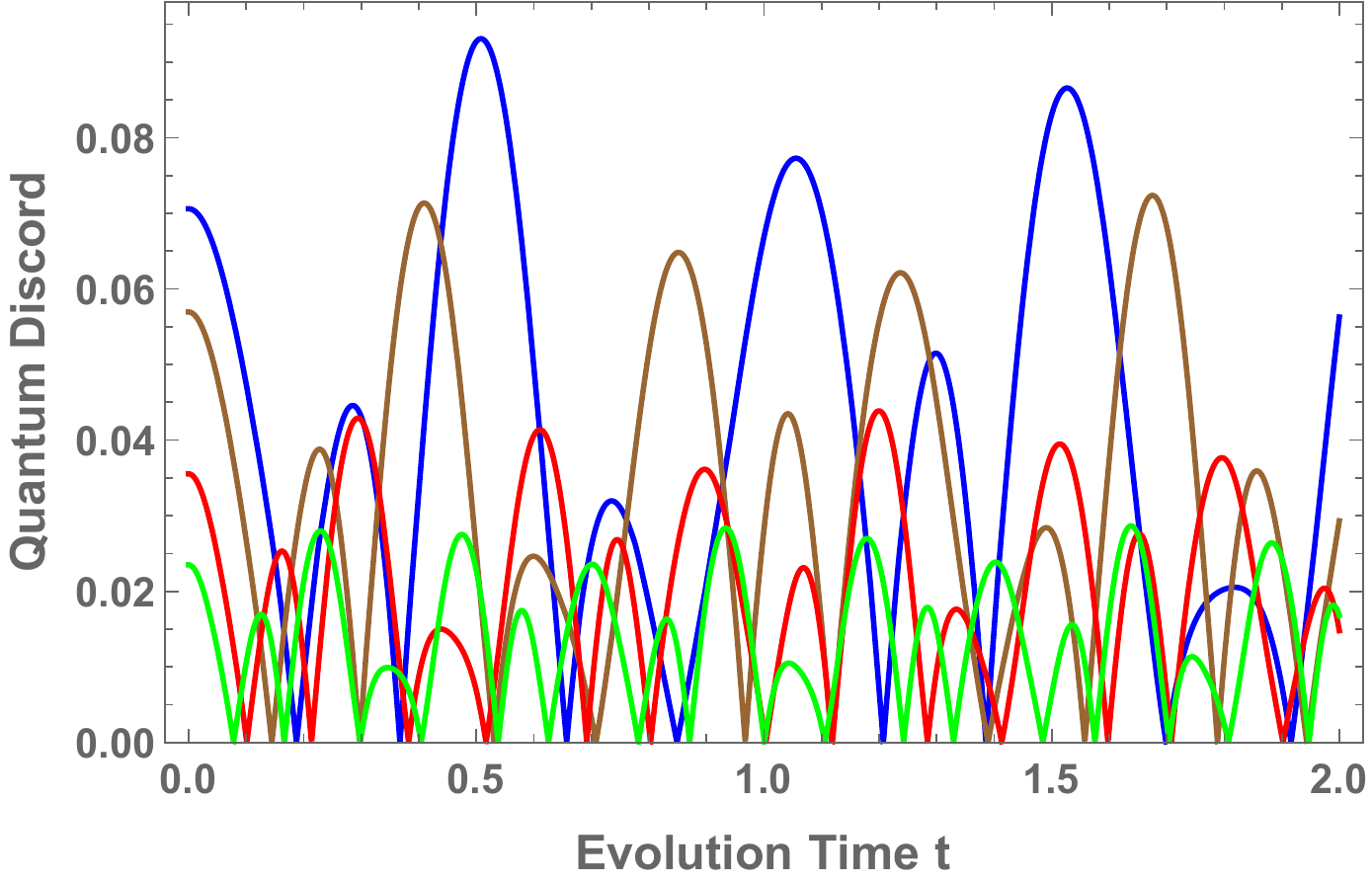} \vfill  $\left(B \right)$
			\end{minipage}\hfill
			\begin{minipage}[b]{.30\linewidth}
				\centering
				\includegraphics[scale=0.42]{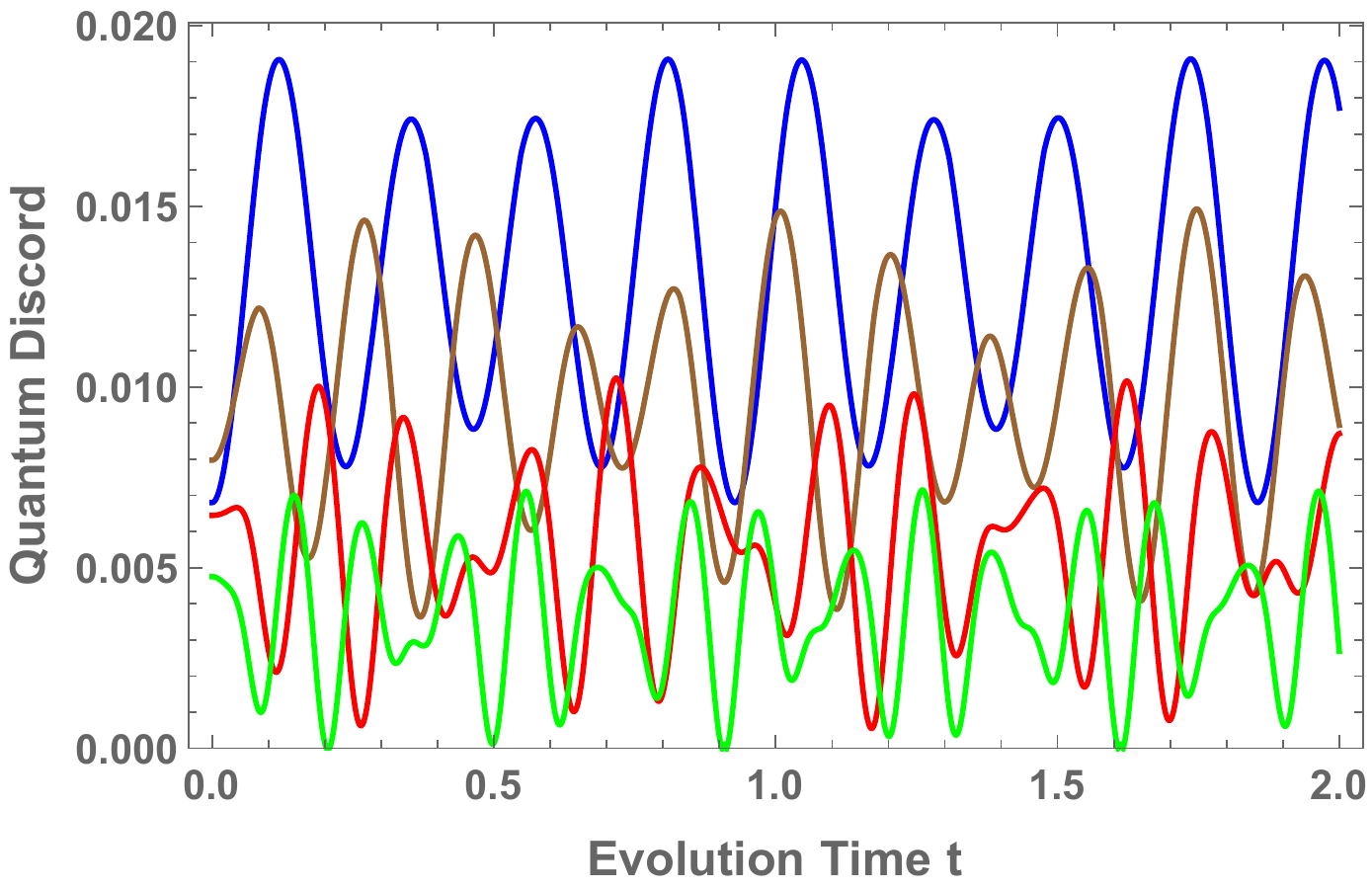} \vfill $\left(C \right)$
	\end{minipage}}}
	\caption{The same as in Figure 3 but for the time evolution of quantum discord in terms of the time $t$}\label{fig4}
\end{figure}
\end{widetext}
The time evolution of the quantum discord for different values of the Stark shift parameters, when the two atoms are initially prepared in their excited state, is depicted in Fig.(\ref{fig3}) for $w_{z}=w_{c}=0$ and in Fig.(\ref{fig4}) when $w_{z}=0.5$ and $w_{c}=1$. We observe from these figures that QD and QC have similar behaviors. As mentioned earlier for QC, the QD evolves periodically with respect to  time and their periods depend heavily on the Stark-shift parameters. Also, increasing the Stark shift parameters leads to reduce the amount of quantum correlations contained in the system. Moreover, QD has confirmed the revival phenomenon like QC and we can find more quantum correlation oscillations for the large values of the Stark shift parameters. We also remark that QD decreases with increasing of photon number as the frequency $w_{z}$ and $w_{c}$ increase. More importantly, by comparing the Figs.\ref{fig1}(A)-(C) with Figs.\ref{fig3}(A)-(C) and Figs.\ref{fig2}(A)-(C) with Figs.\ref{fig4}(A)-(C), we remark that the amount of QC is greater and goes beyond QD. This result is completely in concordance with the physical interpretation given in \cite{Yao2015} where the authors show that QC (via the quantum relative entropy) is more fundamental than other manifestations of quantum correlations (QE and QD) since it can also appear in single-partite systems.

\section{Conclusions}\label{sec:conclusions}
To summarize, we have investigated the quantum coherence dynamics and non-classical correlation dynamics in two two-level atoms interacting with a single-mode electromagnetic cavity field. A special attention is dedicated to the presence of the Stark shifts. Considering two atoms initially in their excited states and the field in coherent state, we have obtained the exact analytic solution of the time-dependent Schrodinger equation and extracted the time evolution of the relevant density matrix elements. In analyzing the dynamics of QC and QD, we observed that the time evolution of quantum coherence quantified by quantum Jensen-Shannon divergence is almost similar to the time evolution of quantum correlations measured by entropic quantum discord. But we have noticed the amount of quantum coherence is always greater and goes beyond quantum discord due to the fact that the total quantum coherence in multipartite systems has contributions from local coherence on subsystems and collective coherence between them. Another remarkable feature for both QC and QD behaviors is the revival phenomenon which reflects the transfer of quantum features from the cavity to the two-qubit system. Furthermore, these two quantifiers decrease with increasing values of the Stark shift parameters as photons number increases. Because of the degeneracy of the levels in the two atoms produced by the electromagnetic cavity field, we remarked that the Stark-shift parameters can dramatically influence both the quantity of the initial quantum correlations and the quantum superposition of the system and can also change the other properties of the physical resources needed for implementing quantum information processing tasks.\par

As prolongation of this work, we believe that it will be interesting to study the effects of the reflecting boundary on the coherence dynamics in this model. Very recently, Huang \cite{Huang2019} has shown that the presence of the boundary in two identical two-level atoms weakly interacting with a bath of fluctuating electromagnetic fields in a vacuum has a significant impact on the generation, revival and degradation of quantum correlations. Furthermore, Cheng et al.\cite{Cheng2018} showed that the boundary executes important influences on the entanglement dynamics behaviors in the same model. On the other hand, this research raised many questions that need further investigation, the most important of which is related to the influence of Stark-shift when the field of photons coupled with a modulated coupling parameter which depends explicitly on time. In other words, our results are encouraging and should be validated in the case of time-dependent fields. We hope to report on this subject in another work.







\begin{thebibliography}{88}%
\makeatletter
\providecommand \@ifxundefined [1]{%
 \@ifx{#1\undefined}
}%
\providecommand \@ifnum [1]{%
 \ifnum #1\expandafter \@firstoftwo
 \else \expandafter \@secondoftwo
 \fi
}%
\providecommand \@ifx [1]{%
 \ifx #1\expandafter \@firstoftwo
 \else \expandafter \@secondoftwo
 \fi
}%
\providecommand \natexlab [1]{#1}%
\providecommand \enquote  [1]{``#1''}%
\providecommand \bibnamefont  [1]{#1}%
\providecommand \bibfnamefont [1]{#1}%
\providecommand \citenamefont [1]{#1}%
\providecommand \href@noop [0]{\@secondoftwo}%
\providecommand \href [0]{\begingroup \@sanitize@url \@href}%
\providecommand \@href[1]{\@@startlink{#1}\@@href}%
\providecommand \@@href[1]{\endgroup#1\@@endlink}%
\providecommand \@sanitize@url [0]{\catcode `\\12\catcode `\$12\catcode
  `\&12\catcode `\#12\catcode `\^12\catcode `\_12\catcode `\%12\relax}%
\providecommand \@@startlink[1]{}%
\providecommand \@@endlink[0]{}%
\providecommand \url  [0]{\begingroup\@sanitize@url \@url }%
\providecommand \@url [1]{\endgroup\@href {#1}{\urlprefix }}%
\providecommand \urlprefix  [0]{URL }%
\providecommand \Eprint [0]{\href }%
\providecommand \doibase [0]{http://dx.doi.org/}%
\providecommand \selectlanguage [0]{\@gobble}%
\providecommand \bibinfo  [0]{\@secondoftwo}%
\providecommand \bibfield  [0]{\@secondoftwo}%
\providecommand \translation [1]{[#1]}%
\providecommand \BibitemOpen [0]{}%
\providecommand \bibitemStop [0]{}%
\providecommand \bibitemNoStop [0]{.\EOS\space}%
\providecommand \EOS [0]{\spacefactor3000\relax}%
\providecommand \BibitemShut  [1]{\csname bibitem#1\endcsname}%
\let\auto@bib@innerbib\@empty
\bibitem [{\citenamefont {Shore1990}\ \emph {et~al.}(1990)\citenamefont {Shore}}]{Shore1990}
\BibitemOpen
\bibfield  {author} {\bibinfo {author} {\bibfnamefont {B.W.}\ \bibnamefont
		{Shore}},\
}\bibfield  {title} {\enquote {\bibinfo {title} {The Theory of Coherent Atomic Excitation},}\ }\href {\doibase
	} {\bibfield  {journal} {\bibinfo  {journal}
		{Wiley, New York, NY, USA}\ }\textbf {\bibinfo {volume} {}},\ \bibinfo {pages}
	{} (\bibinfo {year} {2019})}\BibitemShut {NoStop}
\bibitem [{\citenamefont {Scully1997}\ \emph {et~al.}(1997)\citenamefont {Scully} \ and\ \citenamefont {Zubairy}}]{Scully1997}
\BibitemOpen
\bibfield  {author} {\bibinfo {author} {\bibfnamefont {M.O.}\ \bibnamefont
		{Scully}} \ and\ \bibinfo {author} {\bibfnamefont {M.S.}\ \bibnamefont {Zubairy}},\
}\bibfield  {title} {\enquote {\bibinfo {title} {Quantum Optics},}\ }\href {\doibase
	} {\bibfield  {journal} {\bibinfo  {journal}
		{Cambridge University, Cambridge, England.}\ }\textbf {\bibinfo {volume} {}},\ \bibinfo {pages}
	{} (\bibinfo {year} {1997})}\BibitemShut {NoStop}
\bibitem [{\citenamefont {Allen1997}\ \emph {et~al.}(1997)\citenamefont {Allen} \ and\ \citenamefont {Eberly}}]{Allen1997}
\BibitemOpen
\bibfield  {author} {\bibinfo {author} {\bibfnamefont {L.}\ \bibnamefont
		{Allen}} \ and\ \bibinfo {author} {\bibfnamefont {J.H.}\ \bibnamefont {Eberly}},\
}\bibfield  {title} {\enquote {\bibinfo {title} {Optical Resonance and Two-Level Atoms},}\ }\href {\doibase
	} {\bibfield  {journal} {\bibinfo  {journal}
		{Cambridge University, Cambridge, England.}\ }\textbf {\bibinfo {volume} {}},\ \bibinfo {pages}
	{} (\bibinfo {year} {1997})}\BibitemShut {NoStop}\bibitem [{\citenamefont {Li2019}\ \emph {et~al.}(2019)\citenamefont {Li} \ and\ \citenamefont {Shao}}]{Li2019}
\BibitemOpen
\bibfield  {author} {\bibinfo {author} {\bibfnamefont {D.X.}\ \bibnamefont
		{Li}} \ and\ \bibinfo {author} {\bibfnamefont {X.Q.}\ \bibnamefont {Shao}},\
}\bibfield  {title} {\enquote {\bibinfo {title} {Rapid population transfer of a two-level system by a polychromatic driving field},}\ }\href {\doibase
	10.1038/s41598-019-45558-5} {\bibfield  {journal} {\bibinfo  {journal}
		{Sci. Rep.}\ }\textbf {\bibinfo {volume} {9}},\ \bibinfo {pages}
	{9023} (\bibinfo {year} {2019})}\BibitemShut {NoStop}
\bibitem [{\citenamefont {Nakamura1999}\ \emph {et~al.}(1999)\citenamefont {Nakamura},
	\citenamefont {Pashkin} \ and\ \citenamefont {Tsai}}]{Nakamura1999}
\BibitemOpen
\bibfield  {author} {\bibinfo {author} {\bibfnamefont {Y.}\ \bibnamefont
		{Nakamura}}, \bibinfo {author} {\bibfnamefont {Y.A.}~\bibnamefont
		{Pashkin}}\ and\ \bibinfo {author} {\bibfnamefont {J.S.}\ \bibnamefont {Tsai}},\
}\bibfield  {title} {\enquote {\bibinfo {title} {Coherent control of macroscopic quantum states in a single-Cooper-pair box},}\ }\href {\doibase
	10.1038/19718} {\bibfield  {journal} {\bibinfo  {journal}
		{Nature.}\ }\textbf {\bibinfo {volume} {398}},\ \bibinfo {pages}
	{786–788} (\bibinfo {year} {1999})}\BibitemShut {NoStop}
\bibitem [{\citenamefont {Treutlein2014}\ \emph {et~al.}(2014)\citenamefont {Treutlein},
	\citenamefont {Genes}, \citenamefont {Hammerer}, \citenamefont {Poggio} \ and\ \citenamefont {Rabl}}]{Treutlein2014}
\BibitemOpen
\bibfield  {author} {\bibinfo {author} {\bibfnamefont {P.}\ \bibnamefont
		{Treutlein}}, \bibinfo {author} {\bibfnamefont {C.}~\bibnamefont
		{Genes}}, \bibinfo {author} {\bibfnamefont {K.}~\bibnamefont
		{Hammerer}}, \bibinfo {author} {\bibfnamefont {M.}~\bibnamefont
		{Poggio}}\ and\ \bibinfo {author} {\bibfnamefont {P.}\ \bibnamefont {Rabl}},\
}\bibfield  {title} {\enquote {\bibinfo {title} {Hybrid Mechanical Systems},}\ }\href {\doibase
	10.1007/978-3-642-55312-7\_14} {\bibfield  {journal} {\bibinfo  {journal}
		{Quantum Science and Technology. Springer, Berlin, Heidelberg}\ }\textbf {\bibinfo {volume} {}},\ \bibinfo {pages}
	{327-351} (\bibinfo {year} {2014})}\BibitemShut {NoStop}
\bibitem [{\citenamefont {Lukin2011}\ \emph {et~al.}(2011)\citenamefont {Lukin},
	\citenamefont {Fleischhauer} \ and\ \citenamefont {Imamoğlu}}]{Lukin2011}
\BibitemOpen
\bibfield  {author} {\bibinfo {author} {\bibfnamefont {M.}\ \bibnamefont
		{Lukin}}, \bibinfo {author} {\bibfnamefont {M.}~\bibnamefont
		{Fleischhauer}}\ and\ \bibinfo {author} {\bibfnamefont {A.}\ \bibnamefont {Imamoğlu}},\
}\bibfield  {title} {\enquote {\bibinfo {title} {Quantum information processing based on cavity QED with mesoscopic systems},}\ }\href {\doibase
	} {\bibfield  {journal} {\bibinfo  {journal}
		{Springer  Berlin  Heidelberg.}\ }\textbf {\bibinfo {volume} {}},\ \bibinfo {pages}
	{193-203} (\bibinfo {year} {2011})}\BibitemShut {NoStop}
\bibitem [{\citenamefont {Kang2016}\ \emph {et~al.}(2016)\citenamefont {Kang}, \citenamefont {Chen},\citenamefont {Wu},\citenamefont {Huang},
	\citenamefont {Song} \ and\ \citenamefont {Xia}}]{Kang2016}
\BibitemOpen
\bibfield  {author} {\bibinfo {author} {\bibfnamefont {Y. H.}\ \bibnamefont
		{Kang}}, \bibinfo {author} {\bibfnamefont {Y. H.}\ \bibnamefont
		{Chen}}, \bibinfo {author} {\bibfnamefont {Q. C.}\ \bibnamefont
		{Wu}}, \bibinfo {author} {\bibfnamefont {B. H.}\ \bibnamefont
		{Huang}}, \bibinfo {author} {\bibfnamefont {J.}~\bibnamefont
		{Song}}\ and\ \bibinfo {author} {\bibfnamefont {Y.}\ \bibnamefont {Xia}},\
}\bibfield  {title} {\enquote {\bibinfo {title} {Fast generation of W states of superconducting qubits with multiple Schrödinger dynamics},}\ }\href {\doibase
	10.1038/srep36737} {\bibfield  {journal} {\bibinfo  {journal}
		{Scientific reports.}\ }\textbf {\bibinfo {volume} {6}},\ \bibinfo {pages}
	{36737} (\bibinfo {year} {2016})}\BibitemShut {NoStop}
\bibitem [{\citenamefont {Khan2018}\ \emph {et~al.}(2018)\citenamefont {Khan} \ and\ \citenamefont {Türeci}}]{Khan2018}
\BibitemOpen
\bibfield  {author} {\bibinfo {author} {\bibfnamefont {S.}\ \bibnamefont
		{Khan}}\ and\ \bibinfo {author} {\bibfnamefont {H. E.}\ \bibnamefont {Türeci}},\
}\bibfield  {title} {\enquote {\bibinfo {title} {Frequency combs in a lumped-element josephson-junction circuit},}\ }\href {\doibase
	10.1103/PhysRevLett.120.153601} {\bibfield  {journal} {\bibinfo  {journal}
		{Phys. Rev. Lett.}\ }\textbf {\bibinfo {volume} {120}},\ \bibinfo {pages}
	{153601} (\bibinfo {year} {2018})}\BibitemShut {NoStop}
\bibitem [{\citenamefont {Steane1997}\ \emph {et~al.}(1997)\citenamefont {Steane}}]{Steane1997}
\BibitemOpen
\bibfield  {author} {\bibinfo {author} {\bibfnamefont {A. M.}\ \bibnamefont
		{Steane}},\
}\bibfield  {title} {\enquote {\bibinfo {title} {The ion trap quantum information processor},}\ }\href {\doibase
	10.1007/s003400050225} {\bibfield  {journal} {\bibinfo  {journal}
		{Appl.Phys. B.}\ }\textbf {\bibinfo {volume} {64}},\ \bibinfo {pages}
	{623} (\bibinfo {year} {1997})}\BibitemShut {NoStop}
\bibitem [{\citenamefont {Michler2017}\ \emph {et~al.}(2017)\citenamefont {Michler}}]{Michler2017}
\BibitemOpen
\bibfield  {author} {\bibinfo {author} {\bibfnamefont {P.}\ \bibnamefont
		{Michler}},\
}\bibfield  {title} {\enquote {\bibinfo {title} {Quantum dots for quantum information technologies},}\ }\href {\doibase
} {\bibfield  {journal} {\bibinfo  {journal}
		{Springer international publishing. Berlin}\ }\textbf (\bibinfo {year} {2017})}\BibitemShut {NoStop}
\bibitem [{\citenamefont {Dicke1954}\ \emph {et~al.}(1954)\citenamefont {Dicke}}]{Dicke1954}
\BibitemOpen
\bibfield  {author} {\bibinfo {author} {\bibfnamefont {R. H.}\ \bibnamefont
		{Dicke}},\
}\bibfield  {title} {\enquote {\bibinfo {title} {Coherence in Spontaneous Radiation Processes},}\ }\href {\doibase
10.1103/PhysRev.93.99} {\bibfield  {journal} {\bibinfo  {journal}
	{Phys. Rev.}\ }\textbf {\bibinfo {volume} {93}},\ \bibinfo {pages}
{99} (\bibinfo {year} {1954})}\BibitemShut {NoStop}
\bibitem [{\citenamefont {Abdel-Rady20172}\ \emph {et~al.}(2017)\citenamefont {Abdel-Rady},
	\citenamefont {Hassan}, \citenamefont {Osman} \ and\ \citenamefont {Salah}}]{Abdel-Rady20172}
\BibitemOpen
\bibfield  {author} {\bibinfo {author} {\bibfnamefont {A.S.}\ \bibnamefont
		{Abdel-Rady}}, \bibinfo {author} {\bibfnamefont {S. S.}~\bibnamefont
		{Hassan}}, \bibinfo {author} {\bibfnamefont {A. N. A.}~\bibnamefont
		{Osman}}\ and\ \bibinfo {author} {\bibfnamefont {A.}\ \bibnamefont {Salah}},\
}\bibfield  {title} {\enquote {\bibinfo {title} {Evolution of Extended JC-Dicke Quantum Phase Transition with a Coupled Optical Cavity in Bose-Einstein Condensate System},}\ }\href {\doibase
	10.1007/s10773-017-3531-3} {\bibfield  {journal} {\bibinfo  {journal}
		{Int. J. Theor. Phys}\ }\textbf {\bibinfo {volume} {56}},\ \bibinfo {pages}
	{3655-3666} (\bibinfo {year} {2017})}\BibitemShut {NoStop}
\bibitem [{\citenamefont {Rudolph1998}\ \emph {et~al.}(1998)\citenamefont {Rudolph}, \citenamefont {Freedhoff} \ and\ \citenamefont {Ficek}}]{Rudolph1998}
\BibitemOpen
\bibfield  {author} {\bibinfo {author} {\bibfnamefont {T. G.}\ \bibnamefont
		{Rudolph}}, \bibinfo {author} {\bibfnamefont {H. S.}~\bibnamefont
		{Freedhoff}}\ and\ \bibinfo {author} {\bibfnamefont {Z.}\ \bibnamefont {Ficek}},\
}\bibfield  {title} {\enquote {\bibinfo {title} {Multiphoton ac Stark effect in a bichromatically driven two-level atom},}\ }\href {\doibase
	10.1103/PhysRevA.58.1296} {\bibfield  {journal} {\bibinfo  {journal}
		{Phys. Rev. A}\ }\textbf {\bibinfo {volume} {58}},\ \bibinfo {pages}
	{1296} (\bibinfo {year} {1998})}\BibitemShut {NoStop}
\bibitem [{\citenamefont {Farouk2017}\ \emph {et~al.}(2017)\citenamefont {Obada}, \citenamefont {Ahmed}, \citenamefont {Farouk} \ and\ \citenamefont {Salah}}]{Farouk2017}
\BibitemOpen
\bibfield  {author} {\bibinfo {author} {\bibfnamefont {A. S. F.}\ \bibnamefont
		{Obada}}, \bibinfo {author} {\bibfnamefont {M. M.}~\bibnamefont
		{Ahmed}}, \bibinfo {author} {\bibfnamefont {A. M.}~\bibnamefont
		{Farouk}}\ and\ \bibinfo {author} {\bibfnamefont {A.}\ \bibnamefont {Salah}},\
}\bibfield  {title} {\enquote {\bibinfo {title} {A moving three-level $\Lambda$-type atom in a dissipative cavity},}\ }\href {\doibase
	10.1140/epjd/e2017-80357-5} {\bibfield  {journal} {\bibinfo  {journal}
		{Eur. Phys. J. D.}\ }\textbf {\bibinfo {volume} {71}},\ \bibinfo {pages}
	{338} (\bibinfo {year} {2017})}\BibitemShut {NoStop}	
\bibitem [{\citenamefont {Abdel-Wahab2019}\ \emph {et~al.}(2019)\citenamefont {Abdel-Wahab} \ and\ \citenamefont {Salah}}]{Abdel-Wahab2019}
\BibitemOpen
\bibfield  {author} {\bibinfo {author} {\bibfnamefont {N. H.}\ \bibnamefont
		{Abdel-Wahab}} \ and\ \bibinfo {author} {\bibfnamefont {A.}\ \bibnamefont {Salah}},\
}\bibfield  {title} {\enquote {\bibinfo {title} {On the interaction between a time-dependent field and a two-level atom},}\ }\href {\doibase
	10.1142/S0217732319500810} {\bibfield  {journal} {\bibinfo  {journal}
		{Mod. Phy. Lett. A.}\ }\textbf {\bibinfo {volume} {34}},\ \bibinfo {pages}
	{1950081} (\bibinfo {year} {2019})}\BibitemShut {NoStop}
\bibitem [{\citenamefont {Chen2012}\ \emph {et~al.}(2012)\citenamefont {Chen} \ and\ \citenamefont {Muga}}]{Chen2012}
\BibitemOpen
\bibfield  {author} {\bibinfo {author} {\bibfnamefont {Xi.}\ \bibnamefont
		{Chen}} \ and\ \bibinfo {author} {\bibfnamefont {J. G.}\ \bibnamefont {Muga}},\
}\bibfield  {title} {\enquote {\bibinfo {title} {Engineering of fast population transfer in three-level systems},}\ }\href {\doibase
	10.1103/PhysRevA.86.033405} {\bibfield  {journal} {\bibinfo  {journal}
		{Phys. Rev. A.}\ }\textbf {\bibinfo {volume} {86}},\ \bibinfo {pages}
	{033405} (\bibinfo {year} {2012})}\BibitemShut {NoStop}
\bibitem [{\citenamefont {Baksic2016}\ \emph {et~al.}(2016)\citenamefont {Baksic}, \citenamefont {Ribeiro} \ and\ \citenamefont {Clerk}}]{Baksic2016}
\BibitemOpen
\bibfield  {author} {\bibinfo {author} {\bibfnamefont {A.}\ \bibnamefont
		{Baksic}}, \bibinfo {author} {\bibfnamefont {H.}\ \bibnamefont
		{Ribeiro}} \ and\ \bibinfo {author} {\bibfnamefont {A. A.}\ \bibnamefont {Clerk}},\
}\bibfield  {title} {\enquote {\bibinfo {title} {Speeding up adiabatic quantum state transfer by using dressed states},}\ }\href {\doibase
	10.1103/PhysRevLett.116.230503} {\bibfield  {journal} {\bibinfo  {journal}
		{Phys. Rev. Lett.}\ }\textbf {\bibinfo {volume} {116}},\ \bibinfo {pages}
	{230503} (\bibinfo {year} {2016})}\BibitemShut {NoStop}
\bibitem [{\citenamefont {Chen2017}\ \emph {et~al.}(2017)\citenamefont {Chen}, \citenamefont {Shi}, \citenamefont {Song}, \citenamefont {Xia}\ and\ \citenamefont {Zheng}}]{Chen2017}
\BibitemOpen
\bibfield  {author} {\bibinfo {author} {\bibfnamefont {Y. H.}\ \bibnamefont
		{Chen}}, \bibinfo {author} {\bibfnamefont {Z. C.}\ \bibnamefont
		{Shi}}, \bibinfo {author} {\bibfnamefont {J.}\ \bibnamefont
		{Song}}, \bibinfo {author} {\bibfnamefont {Y.}\ \bibnamefont {Xia}} \ and\ \bibinfo {author} {\bibfnamefont {S. B.}\ \bibnamefont {Zheng}}  ,\
}\bibfield  {title} {\enquote {\bibinfo {title} {Optimal shortcut approach based on an easily obtained intermediate Hamiltonian},}\ }\href {\doibase
	10.1103/PhysRevA.95.062319} {\bibfield  {journal} {\bibinfo  {journal}
		{Phys. Rev. A.}\ }\textbf {\bibinfo {volume} {95}},\ \bibinfo {pages}
	{062319} (\bibinfo {year} {2017})}\BibitemShut {NoStop}
\bibitem [{\citenamefont {Chen2014}\ \emph {et~al.}(2014)\citenamefont {Chen}, \citenamefont {Xia}, \citenamefont {Chen} \ and\ \citenamefont {Song}}]{Chen2014}
\BibitemOpen
\bibfield  {author} {\bibinfo {author} {\bibfnamefont {Y. H.}\ \bibnamefont
		{Chen}}, \bibinfo {author} {\bibfnamefont {Y.}\ \bibnamefont
		{Xia}}, \bibinfo {author} {\bibfnamefont {Q.Q.}\ \bibnamefont
		{Chen}} \ and\ \bibinfo {author} {\bibfnamefont {J.}\ \bibnamefont {Song}},\
}\bibfield  {title} {\enquote {\bibinfo {title} {Efficient shortcuts to adiabatic passage for fast population transfer in multiparticle systems},}\ }\href {\doibase
	10.1103/PhysRevA.89.033856} {\bibfield  {journal} {\bibinfo  {journal}
		{Phys. Rev. A.}\ }\textbf {\bibinfo {volume} {89}},\ \bibinfo {pages}
	{033856} (\bibinfo {year} {2014})}\BibitemShut {NoStop}
\bibitem [{\citenamefont {Chen2015}\ \emph {et~al.}(2015)\citenamefont {Chen}, \citenamefont {Xia}, \citenamefont {Chen} \ and\ \citenamefont {Song}}]{Chen2015}
\BibitemOpen
\bibfield  {author} {\bibinfo {author} {\bibfnamefont {Y. H.}\ \bibnamefont
		{Chen}}, \bibinfo {author} {\bibfnamefont {Y.}\ \bibnamefont
		{Xia}}, \bibinfo {author} {\bibfnamefont {Q.Q.}\ \bibnamefont
		{Chen}} \ and\ \bibinfo {author} {\bibfnamefont {J.}\ \bibnamefont {Song}},\
}\bibfield  {title} {\enquote {\bibinfo {title} {Fast and noise-resistant implementation of quantum phase gates and creation of quantum entangled states},}\ }\href {\doibase
	10.1103/PhysRevA.91.012325} {\bibfield  {journal} {\bibinfo  {journal}
		{Phys. Rev. A.}\ }\textbf {\bibinfo {volume} {91}},\ \bibinfo {pages}
	{012325} (\bibinfo {year} {2015})}\BibitemShut {NoStop}
\bibitem [{\citenamefont {Nielsen2000}\ \emph {et~al.}(2000)\citenamefont {Nielsen}\ and\ \citenamefont {Chuang}}]{Nielsen2000}
\BibitemOpen
\bibfield  {author} {\bibinfo {author} {\bibfnamefont {M. A.}\ \bibnamefont
		{Nielsen}} \ and\ \bibinfo {author} {\bibfnamefont {I. L.}\ \bibnamefont {Chuang}},\
}\bibfield  {title} {\enquote {\bibinfo {title} {Quantum Computation and Quantum Information},}\ }\href {\doibase} {\bibfield  {journal} {\bibinfo  {journal}
		{Cambridge University Press Cambridge.}\ }\textbf (\bibinfo {year} {2000})}\BibitemShut {NoStop}
\bibitem [{\citenamefont {Horodecki2009}\ \emph {et~al.}(2009)\citenamefont {Horodecki},
	\citenamefont {Horodecki}, \citenamefont {Horodecki} \ and\ \citenamefont {Horodecki}}]{Horodecki2009}
\BibitemOpen
\bibfield  {author} {\bibinfo {author} {\bibfnamefont {R.}\ \bibnamefont
		{Horodecki}}, \bibinfo {author} {\bibfnamefont {P.}~\bibnamefont
		{Horodecki}}, \bibinfo {author} {\bibfnamefont {M.}~\bibnamefont
		{Horodecki}} \ and\ \bibinfo {author} {\bibfnamefont {K.}\ \bibnamefont {Horodecki}},\
}\bibfield  {title} {\enquote {\bibinfo {title} {Quantum entanglement},}\ }\href {\doibase
	10.1103/RevModPhys.81.865} {\bibfield  {journal} {\bibinfo  {journal}
		{Rev. Mod. Phys.}\ }\textbf {\bibinfo {volume} {81}},\ \bibinfo {pages}
	{865} (\bibinfo {year} {2009})}\BibitemShut {NoStop}
\bibitem [{\citenamefont {Bennett1993}\ \emph {et~al.}(1993)\citenamefont {Bennett},
	\citenamefont {Brassard}, \citenamefont {Crépeau}, \citenamefont {Jozsa}, \citenamefont {Peres} \ and\ \citenamefont {Wootters}}]{Bennett1993}
\BibitemOpen
\bibfield  {author} {\bibinfo {author} {\bibfnamefont {C. H.}\ \bibnamefont
		{Bennett}}, \bibinfo {author} {\bibfnamefont {G.}~\bibnamefont
		{Brassard}}, \bibinfo {author} {\bibfnamefont {C.}~\bibnamefont
		{Crépeau}}, \bibinfo {author} {\bibfnamefont {R.}~\bibnamefont
		{Jozsa}}, \bibinfo {author} {\bibfnamefont {A.}~\bibnamefont
		{Peres}}\ and\ \bibinfo {author} {\bibfnamefont {W. K.}\ \bibnamefont {Wootters}},\
}\bibfield  {title} {\enquote {\bibinfo {title} {Teleporting an unknown quantum state via dual classical and Einstein-Podolsky-Rosen channels},}\ }\href {\doibase
	10.1103/PhysRevLett.70.1895} {\bibfield  {journal} {\bibinfo  {journal}
		{Phys. Rev. Lett.}\ }\textbf {\bibinfo {volume} {70}},\ \bibinfo {pages}
	{1895-1899} (\bibinfo {year} {1993})}\BibitemShut {NoStop}
\bibitem [{\citenamefont {Bennett1992}\ \emph {et~al.}(1992)\citenamefont {Bennett} \ and\ \citenamefont {Wiesner}}]{Bennett1992}
\BibitemOpen
\bibfield  {author} {\bibinfo {author} {\bibfnamefont {C. H.}\ \bibnamefont
		{Bennett}} \ and\ \bibinfo {author} {\bibfnamefont {S. J.}\ \bibnamefont {Wiesner}},\
}\bibfield  {title} {\enquote {\bibinfo {title} {Communication via one-and two-particle operators on Einstein-Podolsky-Rosen states},}\ }\href {\doibase
	10.1103/PhysRevLett.69.2881} {\bibfield  {journal} {\bibinfo  {journal}
		{Phys. Rev. Lett.}\ }\textbf {\bibinfo {volume} {69}},\ \bibinfo {pages}
	{2881-2884} (\bibinfo {year} {1992})}\BibitemShut {NoStop}
\bibitem [{\citenamefont {Jennewein2000}\ \emph {et~al.}(2000)\citenamefont {Jennewein}, \citenamefont {Simon}, \citenamefont {Weihs}, \citenamefont {Weinfurter} \ and\ \citenamefont {Zeilinger}}]{Jennewein2000}
\BibitemOpen
\bibfield  {author} {\bibinfo {author} {\bibfnamefont {T.}\ \bibnamefont
		{Jennewein}}, \bibinfo {author} {\bibfnamefont {C.}~\bibnamefont
		{Simon}}, \bibinfo {author} {\bibfnamefont {G.}~\bibnamefont
		{Weihs}}, \bibinfo {author} {\bibfnamefont {H.}~\bibnamefont
		{Weinfurter}} \ and\ \bibinfo {author} {\bibfnamefont {A.}\ \bibnamefont {Zeilinger}},\
}\bibfield  {title} {\enquote {\bibinfo {title} {Quantum cryptography with entangled photons},}\ }\href {\doibase
	10.1103/PhysRevLett.84.4729} {\bibfield  {journal} {\bibinfo  {journal}
		{Phys. Rev. Lett.}\ }\textbf {\bibinfo {volume} {84}},\ \bibinfo {pages}
	{4729} (\bibinfo {year} {2000})}\BibitemShut {NoStop}
\bibitem [{\citenamefont {Murao1999}\ \emph {et~al.}(1999)\citenamefont {Murao},
	\citenamefont {Jonathan}, \citenamefont {Plenio} \ and\ \citenamefont {Vedral}}]{Murao1999}
\BibitemOpen
\bibfield  {author} {\bibinfo {author} {\bibfnamefont {M.}\ \bibnamefont
		{Murao}}, \bibinfo {author} {\bibfnamefont {D.}~\bibnamefont
		{Jonathan}}, \bibinfo {author} {\bibfnamefont {M. B.}~\bibnamefont
		{Plenio}}\ and\ \bibinfo {author} {\bibfnamefont {V.}\ \bibnamefont {Vedral}},\
}\bibfield  {title} {\enquote {\bibinfo {title} {Quantum telecloning and multiparticle entanglement},}\ }\href {\doibase
	10.1103/PhysRevA.67.063804} {\bibfield  {journal} {\bibinfo  {journal}
		{Phys. Rev. A.}\ }\textbf {\bibinfo {volume} {59}},\ \bibinfo {pages}
	{156} (\bibinfo {year} {1999})}\BibitemShut {NoStop}
\bibitem [{\citenamefont {Vedral1997}\ \emph {et~al.}(1997)\citenamefont {Vedral},
	\citenamefont {Plenio}, \citenamefont {Rippin} \ and\ \citenamefont {Knight}}]{Vedral1997}
\BibitemOpen
\bibfield  {author} {\bibinfo {author} {\bibfnamefont {V.}\ \bibnamefont
		{Vedral}}, \bibinfo {author} {\bibfnamefont {M. B.}~\bibnamefont
		{Plenio}}, \bibinfo {author} {\bibfnamefont {M. A.}~\bibnamefont
		{Rippin}}\ and\ \bibinfo {author} {\bibfnamefont {P. L.}\ \bibnamefont {Knight}},\
}\bibfield  {title} {\enquote {\bibinfo {title} {Quantifying entanglement},}\ }\href {\doibase
	10.1103/PhysRevLett.78.2275} {\bibfield  {journal} {\bibinfo  {journal}
		{Phys. Rev. Lett.}\ }\textbf {\bibinfo {volume} {78}},\ \bibinfo {pages}
	{2275} (\bibinfo {year} {1997})}\BibitemShut {NoStop}
\bibitem [{\citenamefont {Braun2002}\ \emph {et~al.}(2002)\citenamefont {Braun}}]{Braun2002}
\BibitemOpen
\bibfield  {author} {\bibinfo {author} {\bibfnamefont {D.}\ \bibnamefont
		{Braun}},\
}\bibfield  {title} {\enquote {\bibinfo {title} {Creation of entanglement by interaction with a common heat bath},}\ }\href {\doibase
	10.1103/PhysRevLett.89.277901} {\bibfield  {journal} {\bibinfo  {journal}
		{Phys. Rev. Lett.}\ }\textbf {\bibinfo {volume} {89}},\ \bibinfo {pages}
	{277901} (\bibinfo {year} {2002})}\BibitemShut {NoStop}
\bibitem [{\citenamefont {Slaoui20182}\ \emph {et~al.}(2018)\citenamefont {Slaoui},
	\citenamefont {Shaukat}, \citenamefont {Daoud} \ and\ \citenamefont {Laamara}}]{Slaoui20182}
\BibitemOpen
\bibfield  {author} {\bibinfo {author} {\bibfnamefont {A.}\ \bibnamefont
		{Slaoui}}, \bibinfo {author} {\bibfnamefont {M. I.}~\bibnamefont
		{Shaukat}}, \bibinfo {author} {\bibfnamefont {M.}~\bibnamefont
		{Daoud}}\ and\ \bibinfo {author} {\bibfnamefont {R. A.}\ \bibnamefont {Laamara}},\
}\bibfield  {title} {\enquote {\bibinfo {title} {Universal evolution of non-classical correlations due to collective spontaneous emission},}\ }\href {\doibase
	10.1140/epjp/i2018-12211-y} {\bibfield  {journal} {\bibinfo  {journal}
		{Eur. Phys. J. Plus.}\ }\textbf {\bibinfo {volume} {133}},\ \bibinfo {pages}
	{413} (\bibinfo {year} {2018})}\BibitemShut {NoStop}
\bibitem [{\citenamefont {Yu2004}\ \emph {et~al.}(2004)\citenamefont {Yu} \ and\ \citenamefont {Eberly}}]{Yu2004}
\BibitemOpen
\bibfield  {author} {\bibinfo {author} {\bibfnamefont {T.}\ \bibnamefont
		{Yu}}\ and\ \bibinfo {author} {\bibfnamefont {J.  H.}\ \bibnamefont {Eberly}},\
}\bibfield  {title} {\enquote {\bibinfo {title} {Finite-time disentanglement via spontaneous emission},}\ }\href {\doibase
	10.1103/PhysRevLett.93.140404} {\bibfield  {journal} {\bibinfo  {journal}
		{Phys. Rev. Lett.}\ }\textbf {\bibinfo {volume} {93}},\ \bibinfo {pages}
	{140404} (\bibinfo {year} {2004})}\BibitemShut {NoStop}
\bibitem [{\citenamefont {Datta2005}\ \emph {et~al.}(2005)\citenamefont {Datta},
	\citenamefont {Flammia} \ and\ \citenamefont {Caves}}]{Datta2005}
\BibitemOpen
\bibfield  {author} {\bibinfo {author} {\bibfnamefont {A.}\ \bibnamefont
		{Datta}}, \bibinfo {author} {\bibfnamefont {A. T.}~\bibnamefont
		{Flammia}}\ and\ \bibinfo {author} {\bibfnamefont {C. M.}\ \bibnamefont {Caves}},\
}\bibfield  {title} {\enquote {\bibinfo {title} {Entanglement and the power of one qubit},}\ }\href {\doibase
	10.1103/PhysRevA.72.042316} {\bibfield  {journal} {\bibinfo  {journal}
		{Phys. Rev. A.}\ }\textbf {\bibinfo {volume} {72}},\ \bibinfo {pages}
	{042316} (\bibinfo {year} {2005})}\BibitemShut {NoStop}
\bibitem [{\citenamefont {Datta2007}\ \emph {et~al.}(2007)\citenamefont {Datta} \ and\ \citenamefont {Vidal}}]{Datta2007}
\BibitemOpen
\bibfield  {author} {\bibinfo {author} {\bibfnamefont {A.}\ \bibnamefont
		{Datta}}\ and\ \bibinfo {author} {\bibfnamefont {G.}\ \bibnamefont {Vidal}},\
}\bibfield  {title} {\enquote {\bibinfo {title} {Role of entanglement and correlations in mixed-state quantum computation},}\ }\href {\doibase
	10.1103/PhysRevA.75.042310} {\bibfield  {journal} {\bibinfo  {journal}
		{Phys. Rev. A.}\ }\textbf {\bibinfo {volume} {75}},\ \bibinfo {pages}
	{042310} (\bibinfo {year} {2007})}\BibitemShut {NoStop}
\bibitem [{\citenamefont {Braunstein1999}\ \emph {et~al.}(1999)\citenamefont {Braunstein},
	\citenamefont {Caves}, \citenamefont {Jozsa},\citenamefont {Linden}, \citenamefont {Popescu} \ and\ \citenamefont {Schack}}]{Braunstein1999}
\BibitemOpen
\bibfield  {author} {\bibinfo {author} {\bibfnamefont {S. L.}\ \bibnamefont
		{Braunstein}},\bibinfo {author} {\bibfnamefont {C. M.}~\bibnamefont
		{Caves}},\bibinfo {author} {\bibfnamefont {R. }~\bibnamefont
		{Jozsa}},\bibinfo {author} {\bibfnamefont {N. }~\bibnamefont
		{Linden}},\bibinfo {author} {\bibfnamefont {S. }~\bibnamefont
		{Popescu}}\ and\ \bibinfo {author} {\bibfnamefont {R.}\ \bibnamefont {Schack}},\
}\bibfield  {title} {\enquote {\bibinfo {title} {Separability of very noisy mixed states and implications for NMR quantum computing},}\ }\href {\doibase
	10.1103/PhysRevLett.83.1054} {\bibfield  {journal} {\bibinfo  {journal}
		{Phys. Rev. Lett.}\ }\textbf {\bibinfo {volume} {83}},\ \bibinfo {pages}
	{1054} (\bibinfo {year} {1999})}\BibitemShut {NoStop}
\bibitem [{\citenamefont {Meyer2000}\ \emph {et~al.}(2000)\citenamefont {Meyer}}]{Meyer2000}
\BibitemOpen
\bibfield  {author} {\bibinfo {author} {\bibfnamefont {D. A.}\ \bibnamefont
		{Meyer}},\
}\bibfield  {title} {\enquote {\bibinfo {title} {Sophisticated quantum search without entanglement},}\ }\href {\doibase
	10.1103/PhysRevLett.85.2014} {\bibfield  {journal} {\bibinfo  {journal}
		{Phys. Rev. Lett.}\ }\textbf {\bibinfo {volume} {85}},\ \bibinfo {pages}
	{2014} (\bibinfo {year} {2000})}\BibitemShut {NoStop}
\bibitem [{\citenamefont {Bennett1999}\ \emph {et~al.}(1999)\citenamefont {Bennett},
	\citenamefont {DiVincenzo}, \citenamefont {Fuchs}, \citenamefont {Mor}, \citenamefont {Rains}, \citenamefont {Shor}, \citenamefont {Smolin} \ and\ \citenamefont {Wootters}}]{Bennett1999}
\BibitemOpen
\bibfield  {author} {\bibinfo {author} {\bibfnamefont {C. H.}\ \bibnamefont
		{Bennett}}, \bibinfo {author} {\bibfnamefont {D. P.}~\bibnamefont
		{DiVincenzo}}, \bibinfo {author} {\bibfnamefont {C. A.}~\bibnamefont
		{Fuchs}}, \bibinfo {author} {\bibfnamefont {T.}~\bibnamefont
		{Mor}}, \bibinfo {author} {\bibfnamefont {E.}~\bibnamefont
		{Rains}}, \bibinfo {author} {\bibfnamefont {P. W.}~\bibnamefont
		{Shor}}, \bibinfo {author} {\bibfnamefont {J. A.}~\bibnamefont
		{Smolin}}, \ and\ \bibinfo {author} {\bibfnamefont {W. K.}\ \bibnamefont {Wootters}},\
}\bibfield  {title} {\enquote {\bibinfo {title} {Quantum nonlocality without entanglement},}\ }\href {\doibase
	10.1103/PhysRevA.59.1070} {\bibfield  {journal} {\bibinfo  {journal}
		{Phys. Rev. A.}\ }\textbf {\bibinfo {volume} {59}},\ \bibinfo {pages}
	{1070} (\bibinfo {year} {1999})}\BibitemShut {NoStop}\bibitem [{\citenamefont {Niset2006}\ \emph {et~al.}(2006)\citenamefont {Niset}\ and\ \citenamefont {Cerf}}]{Niset2006}
\BibitemOpen
\bibfield  {author} {\bibinfo {author} {\bibfnamefont {J.}\ \bibnamefont
		{Niset}}\ and\ \bibinfo {author} {\bibfnamefont {N. J.}\ \bibnamefont {Cerf}},\
}\bibfield  {title} {\enquote {\bibinfo {title} {Multipartite nonlocality without entanglement in many dimensions},}\ }\href {\doibase
	10.1103/PhysRevA.74.052103} {\bibfield  {journal} {\bibinfo  {journal}
		{Phys. Rev. A.}\ }\textbf {\bibinfo {volume} {74}},\ \bibinfo {pages}
	{052103} (\bibinfo {year} {2006})}\BibitemShut {NoStop}\bibitem [{\citenamefont {Lanyon2008}\ \emph {et~al.}(2008)\citenamefont {Lanyon},
	\citenamefont {Barbieri}, \citenamefont {Almeida}\ and\ \citenamefont {White}}]{Lanyon2008}
\BibitemOpen
\bibfield  {author} {\bibinfo {author} {\bibfnamefont {B. P.}\ \bibnamefont
		{Lanyon}}, \bibinfo {author} {\bibfnamefont {M.}~\bibnamefont
		{Barbieri}}, \bibinfo {author} {\bibfnamefont {M. P.}~\bibnamefont
		{Almeida}}\ and\ \bibinfo {author} {\bibfnamefont {A. G.}\ \bibnamefont {White}},\
}\bibfield  {title} {\enquote {\bibinfo {title} {Experimental quantum computing without entanglement},}\ }\href {\doibase
	10.1103/PhysRevLett.101.200501} {\bibfield  {journal} {\bibinfo  {journal}
		{Phys. Rev. Lett.}\ }\textbf {\bibinfo {volume} {101}},\ \bibinfo {pages}
	{200501} (\bibinfo {year} {2008})}\BibitemShut {NoStop}
\bibitem [{\citenamefont {Ollivier2001}\ \emph {et~al.}(2001)\citenamefont {Ollivier} \ and\ \citenamefont {Zurek}}]{Ollivier2001}
\BibitemOpen
\bibfield  {author} {\bibinfo {author} {\bibfnamefont {H.}\ \bibnamefont
		{Ollivier}}\ and\ \bibinfo {author} {\bibfnamefont {W. H.}\ \bibnamefont {Zurek}},\
}\bibfield  {title} {\enquote {\bibinfo {title} {Quantum discord: a measure of the quantumness of correlations},}\ }\href {\doibase
	10.1103/PhysRevLett.88.017901} {\bibfield  {journal} {\bibinfo  {journal}
		{Phys. Rev. Lett.}\ }\textbf {\bibinfo {volume} {88}},\ \bibinfo {pages}
	{017901} (\bibinfo {year} {2001})}\BibitemShut {NoStop}
\bibitem [{\citenamefont {Henderson2001}\ \emph {et~al.}(2001)\citenamefont {Henderson} \ and\ \citenamefont {Vedral}}]{Henderson2001}
\BibitemOpen
\bibfield  {author} {\bibinfo {author} {\bibfnamefont {L.}\ \bibnamefont
		{Henderson}} \ and\ \bibinfo {author} {\bibfnamefont {V.}\ \bibnamefont {Vedral}},\
}\bibfield  {title} {\enquote {\bibinfo {title} {Classical, quantum and total correlations},}\ }\href {\doibase
	10.1088/0305-4470/34/35/315} {\bibfield  {journal} {\bibinfo  {journal}
		{J. Phys A: Math. General}\ }\textbf {\bibinfo {volume} {34}},\ \bibinfo {pages}
	{6899} (\bibinfo {year} {2001})}\BibitemShut {NoStop}
\bibitem [{\citenamefont {Fanchini2011}\ \emph {et~al.}(2011)\citenamefont {Fanchini}, \citenamefont {Cornelio}, \citenamefont {de Oliveira} \ and\ \citenamefont {Caldeira}}]{Fanchini2011}
\BibitemOpen
\bibfield  {author} {\bibinfo {author} {\bibfnamefont {F. F.}\ \bibnamefont
		{Fanchini}}, \bibinfo {author} {\bibfnamefont {M. F.}\ \bibnamefont
		{Cornelio}}, \bibinfo {author} {\bibfnamefont {M. C.}\ \bibnamefont
		{de Oliveira}} \ and\ \bibinfo {author} {\bibfnamefont {A. O.}\ \bibnamefont {Caldeira}},\
}\bibfield  {title} {\enquote {\bibinfo {title} {Conservation law for distributed entanglement of formation and quantum discord},}\ }\href {\doibase
	10.1103/PhysRevA.84.012313} {\bibfield  {journal} {\bibinfo  {journal}
		{Phys. Rev. A.}\ }\textbf {\bibinfo {volume} {84}},\ \bibinfo {pages}
	{012313} (\bibinfo {year} {2011})}\BibitemShut {NoStop}
\bibitem [{\citenamefont {Cornelio2011}\ \emph {et~al.}(2011)\citenamefont {Cornelio}, \citenamefont {de Oliveira} \ and\ \citenamefont {Fanchini}}]{Cornelio2011}
\BibitemOpen
\bibfield  {author} {\bibinfo {author} {\bibfnamefont {M. F.}\ \bibnamefont
		{Cornelio}}, \bibinfo {author} {\bibfnamefont {M. C.}\ \bibnamefont
		{de Oliveira}} \ and\ \bibinfo {author} {\bibfnamefont {F. F.}\ \bibnamefont {Fanchini}},\
}\bibfield  {title} {\enquote {\bibinfo {title} {Entanglement irreversibility from quantum discord and quantum deficit},}\ }\href {\doibase
	10.1103/PhysRevLett.107.020502} {\bibfield  {journal} {\bibinfo  {journal}
		{Phys. Rev. Lett.}\ }\textbf {\bibinfo {volume} {107}},\ \bibinfo {pages}
	{020502} (\bibinfo {year} {2011})}\BibitemShut {NoStop}
\bibitem [{\citenamefont {Dakic2012}\ \emph {et~al.}(2012)\citenamefont {Dakić}}]{Dakic2012}
\BibitemOpen
\bibfield  {author} {\bibinfo {author} {\bibfnamefont {B.}\ \bibnamefont
		{Dakić et~al.}}\
}\bibfield  {title} {\enquote {\bibinfo {title} {Quantum discord as resource for remote state preparation},}\ }\href {\doibase
	10.1038/nphys2377} {\bibfield  {journal} {\bibinfo  {journal}
		{Nat. Phys.}\ }\textbf {\bibinfo {volume} {8}},\ \bibinfo {pages}
	{666-670} (\bibinfo {year} {2012})}\BibitemShut {NoStop}
\bibitem [{\citenamefont {Datta2008}\ \emph {et~al.}(2008)\citenamefont {Datta}, \citenamefont {Shaji} \ and\ \citenamefont {Synak-Radtke}}]{Datta2008}
\BibitemOpen
\bibfield  {author} {\bibinfo {author} {\bibfnamefont {A.}\ \bibnamefont
		{Datta}}, \bibinfo {author} {\bibfnamefont {A.}~\bibnamefont
		{Shaji}}\ and\ \bibinfo {author} {\bibfnamefont {C. M.}\ \bibnamefont {Caves}},\
}\bibfield  {title} {\enquote {\bibinfo {title} {Quantum discord and the power of one qubit},}\ }\href {\doibase
	10.1103/PhysRevLett.100.050502} {\bibfield  {journal} {\bibinfo  {journal}
		{Phys. Rev. Lett.}\ }\textbf {\bibinfo {volume} {100}},\ \bibinfo {pages}
	{050502} (\bibinfo {year} {2008})}\BibitemShut {NoStop}
\bibitem [{\citenamefont {Werlang2009}\ \emph {et~al.}(2009)\citenamefont {Werlang},
	\citenamefont {Souza}, \citenamefont {Fanchini}\ and\ \citenamefont {Boas}}]{Werlang2009}
\BibitemOpen
\bibfield  {author} {\bibinfo {author} {\bibfnamefont {T.}\ \bibnamefont
		{Werlang}}, \bibinfo {author} {\bibfnamefont {S.}~\bibnamefont
		{Souza}}, \bibinfo {author} {\bibfnamefont {F. F.}~\bibnamefont
		{Fanchini}}\ and\ \bibinfo {author} {\bibfnamefont {C. V.}\ \bibnamefont {Boas}},\
}\bibfield  {title} {\enquote {\bibinfo {title} {Robustness of quantum discord to sudden death},}\ }\href {\doibase
	10.1103/PhysRevA.80.024103} {\bibfield  {journal} {\bibinfo  {journal}
		{Phys. Rev. A.}\ }\textbf {\bibinfo {volume} {80}},\ \bibinfo {pages}
	{024103} (\bibinfo {year} {2009})}\BibitemShut {NoStop}
\bibitem [{\citenamefont {Horodecki2005}\ \emph {et~al.}(2005)\citenamefont {Horodecki},
	\citenamefont {Horodecki}, \citenamefont {Horodecki}, \citenamefont {Oppenheim}, \citenamefont {Sen}, \citenamefont {Sen} \ and\ \citenamefont {Synak-Radtke}}]{Horodecki2005}
\BibitemOpen
\bibfield  {author} {\bibinfo {author} {\bibfnamefont {M.}\ \bibnamefont
		{Horodecki}}, \bibinfo {author} {\bibfnamefont {P.}~\bibnamefont
		{Horodecki}}, \bibinfo {author} {\bibfnamefont {R.}~\bibnamefont
		{Horodecki}}, \bibinfo {author} {\bibfnamefont {J.}~\bibnamefont
		{Oppenheim}}, \bibinfo {author} {\bibfnamefont {A.}~\bibnamefont
		{Sen}}, \bibinfo {author} {\bibfnamefont {U.}~\bibnamefont
		{Sen}} \ and\ \bibinfo {author} {\bibfnamefont {B.}\ \bibnamefont {Synak-Radtke}},\
}\bibfield  {title} {\enquote {\bibinfo {title} {Local versus nonlocal information in quantum-information theory: formalism and phenomena},}\ }\href {\doibase
	10.1103/PhysRevA.71.062307} {\bibfield  {journal} {\bibinfo  {journal}
		{Phys. Rev. A.}\ }\textbf {\bibinfo {volume} {71}},\ \bibinfo {pages}
	{062307} (\bibinfo {year} {2005})}\BibitemShut {NoStop}
\bibitem [{\citenamefont {Girolami2013}\ \emph {et~al.}(2013)\citenamefont {Girolami},
	\citenamefont {Tufarelli} \ and\ \citenamefont {Adesso}}]{Girolami2013}
\BibitemOpen
\bibfield  {author} {\bibinfo {author} {\bibfnamefont {D.}\ \bibnamefont
		{Girolami}}, \bibinfo {author} {\bibfnamefont {T.}~\bibnamefont
		{Tufarelli}}\ and\ \bibinfo {author} {\bibfnamefont {G.}\ \bibnamefont {Adesso}},\
}\bibfield  {title} {\enquote {\bibinfo {title} {Characterizing nonclassical correlations via local quantum uncertainty},}\ }\href {\doibase
	10.1103/PhysRevLett.110.240402} {\bibfield  {journal} {\bibinfo  {journal}
		{Phys. Rev. Lett.}\ }\textbf {\bibinfo {volume} {110}},\ \bibinfo {pages}
	{240402} (\bibinfo {year} {2013})}\BibitemShut {NoStop}
\bibitem [{\citenamefont {Slaoui2018}\ \emph {et~al.}(2018)\citenamefont {Slaoui},
	\citenamefont {Daoud} \ and\ \citenamefont {Laamara}}]{Slaoui2018}
\BibitemOpen
\bibfield  {author} {\bibinfo {author} {\bibfnamefont {A.}\ \bibnamefont
		{Slaoui}}, \bibinfo {author} {\bibfnamefont {M.}~\bibnamefont
		{Daoud}}\ and\ \bibinfo {author} {\bibfnamefont {R. A.}\ \bibnamefont {Laamara}},\
}\bibfield  {title} {\enquote {\bibinfo {title} {The dynamics of local quantum uncertainty and trace distance discord for two-qubit X-states under decoherence: a comparative study},}\ }\href {\doibase
	10.1007/s11128-018-1942-6} {\bibfield  {journal} {\bibinfo  {journal}
		{Quantum Inf. Process.}\ }\textbf {\bibinfo {volume} {17}},\ \bibinfo {pages}
	{178} (\bibinfo {year} {2018})}\BibitemShut {NoStop}
\bibitem [{\citenamefont {Slaoui2019}\ \emph {et~al.}(2019)\citenamefont {Slaoui},
	\citenamefont {Daoud} \ and\ \citenamefont {Laamara}}]{Slaoui2019}
\BibitemOpen
\bibfield  {author} {\bibinfo {author} {\bibfnamefont {A.}\ \bibnamefont
		{Slaoui}}, \bibinfo {author} {\bibfnamefont {M.}~\bibnamefont
		{Daoud}}\ and\ \bibinfo {author} {\bibfnamefont {R. A.}\ \bibnamefont {Laamara}},\
}\bibfield  {title} {\enquote {\bibinfo {title} {The dynamic behaviors of local quantum uncertainty for three-qubit $X$ states under decoherence channels},}\ }\href {\doibase
	10.1007/s11128-019-2363-x} {\bibfield  {journal} {\bibinfo  {journal}
		{Quantum Inf. Process.}\ }\textbf {\bibinfo {volume} {18}},\ \bibinfo {pages}
	{250} (\bibinfo {year} {2019})}\BibitemShut {NoStop}
\bibitem [{\citenamefont {Kim2018}\ \emph {et~al.}(2018)\citenamefont {Kim},
	\citenamefont {Li}, \citenamefont {Kumar} \ and\ \citenamefont {Wu}}]{Kim2018}
\BibitemOpen
\bibfield  {author} {\bibinfo {author} {\bibfnamefont {S.}\ \bibnamefont
		{Kim}}, \bibinfo {author} {\bibfnamefont {L.}~\bibnamefont
		{Li}}, \bibinfo {author} {\bibfnamefont {A.}~\bibnamefont
		{Kumar}}\ and\ \bibinfo {author} {\bibfnamefont {J.}\ \bibnamefont {Wu}},\
}\bibfield  {title} {\enquote {\bibinfo {title} {Characterizing nonclassical correlations via local quantum Fisher information},}\ }\href {\doibase
	10.1103/PhysRevA.97.032326} {\bibfield  {journal} {\bibinfo  {journal}
		{Phys. Rev. A.}\ }\textbf {\bibinfo {volume} {97}},\ \bibinfo {pages}
	{032326} (\bibinfo {year} {2018})}\BibitemShut {NoStop}
\bibitem [{\citenamefont {Slaoui20192}\ \emph {et~al.}(2019)\citenamefont {Slaoui},
	\citenamefont {Bakmou}, \citenamefont {Daoud} \ and\ \citenamefont {Laamara}}]{Slaoui20192}
\BibitemOpen
\bibfield  {author} {\bibinfo {author} {\bibfnamefont {A.}\ \bibnamefont
		{Slaoui}}, \bibinfo {author} {\bibfnamefont {L.}~\bibnamefont
		{Bakmou}}, \bibinfo {author} {\bibfnamefont {M.}~\bibnamefont
		{Daoud}}\ and\ \bibinfo {author} {\bibfnamefont {R. A.}\ \bibnamefont {Laamara}},\
}\bibfield  {title} {\enquote {\bibinfo {title} {A comparative study of local quantum Fisher information and local quantum uncertainty in Heisenberg $XY$ model},}\ }\href {\doibase
	10.1016/j.physleta.2019.04.040} {\bibfield  {journal} {\bibinfo  {journal}
		{Phys. Lett. A.}\ }\textbf {\bibinfo {volume} {383}},\ \bibinfo {pages}
	{2241-2247} (\bibinfo {year} {2019})}\BibitemShut {NoStop}
\bibitem [{\citenamefont {Girolami2014}\ \emph {et~al.}(2014)\citenamefont {Girolami},
	\citenamefont {Souza}, \citenamefont {Giovannetti}, \citenamefont {Tufarelli}, \citenamefont {Filgueiras}, \citenamefont {Sarthour}, \citenamefont {Soares-Pinto}, \citenamefont {Oliveira} \ and\ \citenamefont {Adesso}}]{Girolami2014}
\BibitemOpen
\bibfield  {author} {\bibinfo {author} {\bibfnamefont {D.}\ \bibnamefont
		{Girolami}}, \bibinfo {author} {\bibfnamefont {A. M.}~\bibnamefont
		{Souza}}, \bibinfo {author} {\bibfnamefont {V.}~\bibnamefont
		{Giovannetti}}, \bibinfo {author} {\bibfnamefont {T.}~\bibnamefont
		{Tufarelli}}, \bibinfo {author} {\bibfnamefont {J. G.}~\bibnamefont
		{Filgueiras}}, \bibinfo {author} {\bibfnamefont {R. S.}~\bibnamefont
		{Sarthour}}, \bibinfo {author} {\bibfnamefont {D. O.}~\bibnamefont
		{Soares-Pinto}}, \bibinfo {author} {\bibfnamefont {I. S.}~\bibnamefont
		{Oliveira}}\ and\ \bibinfo {author} {\bibfnamefont {G.}\ \bibnamefont {Adesso}},\
}\bibfield  {title} {\enquote {\bibinfo {title} {Quantum discord determines the interferometric power of quantum states},}\ }\href {\doibase
	10.1103/PhysRevLett.112.210401} {\bibfield  {journal} {\bibinfo  {journal}
		{Phys. Rev. Lett.}\ }\textbf {\bibinfo {volume} {112}},\ \bibinfo {pages}
	{210401} (\bibinfo {year} {2014})}\BibitemShut {NoStop}
\bibitem [{\citenamefont {Zurek2003}\ \emph {et~al.}(2003)\citenamefont {Zurek}}]{Zurek2003}
\BibitemOpen
\bibfield  {author} {\bibinfo {author} {\bibfnamefont {W. H.}\ \bibnamefont
		{Zurek}},\
}\bibfield  {title} {\enquote {\bibinfo {title} {Decoherence, einselection, and the quantum origins of the classical},}\ }\href {\doibase
	10.1103/RevModPhys.75.715} {\bibfield  {journal} {\bibinfo  {journal}
		{Rev. Mod. Phys.}\ }\textbf {\bibinfo {volume} {75}},\ \bibinfo {pages}
	{715} (\bibinfo {year} {2003})}\BibitemShut {NoStop}
\bibitem [{\citenamefont {Aberg2006}\ \emph {}(2006)\citenamefont {Aberg}}]{Aberg2006}
\BibitemOpen
\bibfield  {author} {\bibinfo {author} {\bibfnamefont {J.}\ \bibnamefont
		{Åberg}},\
}\bibfield  {title} {\enquote {\bibinfo {title} {Quantifying superposition},}\ }\href {\doibase
	arXiv preprint (2006) arXiv:quant-ph/﻿0612146 [quant-ph]} {\bibfield  {journal} {\bibinfo  {journal}
		{arXiv preprint (2006) arXiv:quant-ph/﻿0612146 [quant-ph]}\ }\ }\BibitemShut {NoStop}
\bibitem [{\citenamefont {Aberg2014}\ \emph {et~al.}(2014)\citenamefont {Åberg} }]{Aberg2014}
\BibitemOpen
\bibfield  {author} {\bibinfo {author} {\bibfnamefont {J.}\ \bibnamefont {Åberg}},\
}\bibfield  {title} {\enquote {\bibinfo {title} {Catalytic Coherence},}\ }\href {\doibase
	10.1103/PhysRevLett.113.150402} {\bibfield  {journal} {\bibinfo  {journal}
		{Phys. Rev. Lett.}\ }\textbf {\bibinfo {volume} {113}},\ \bibinfo {pages}
	{150402} (\bibinfo {year} {2014})}\BibitemShut {NoStop}
\bibitem [{\citenamefont {Baumgratz2014}\ \emph {et~al.}(2014)\citenamefont {Baumgratz},
	\citenamefont {Cramer} \ and\ \citenamefont {Plenio}}]{Baumgratz2014}
\BibitemOpen
\bibfield  {author} {\bibinfo {author} {\bibfnamefont {T.}\ \bibnamefont
		{Baumgratz}}, \bibinfo {author} {\bibfnamefont {M.}~\bibnamefont
		{Cramer}} \ and\ \bibinfo {author} {\bibfnamefont {M. B.}\ \bibnamefont {Plenio}},\
}\bibfield  {title} {\enquote {\bibinfo {title} {Quantifying Coherence},}\ }\href {\doibase
	10.1103/PhysRevLett.113.140401} {\bibfield  {journal} {\bibinfo  {journal}
		{Phys. Rev. Lett.}\ }\textbf {\bibinfo {volume} {113}},\ \bibinfo {pages}
	{140401} (\bibinfo {year} {2014})}\BibitemShut {NoStop}
\bibitem [{\citenamefont {Streltsov2016}\ \emph {et~al.}(2016)\citenamefont {Streltsov},
	\citenamefont {Chitambar}, \citenamefont {Rana}, \citenamefont {Bera}, \citenamefont {Winter} \ and\ \citenamefont {Lewenstein}}]{Streltsov2016}
\BibitemOpen
\bibfield  {author} {\bibinfo {author} {\bibfnamefont {A.}\ \bibnamefont
		{Streltsov}}, \bibinfo {author} {\bibfnamefont {E.}~\bibnamefont
		{Chitambar}}, \bibinfo {author} {\bibfnamefont {S.}~\bibnamefont
		{Rana}}, \bibinfo {author} {\bibfnamefont {M. N.}~\bibnamefont
		{Bera}}, \bibinfo {author} {\bibfnamefont {A.}~\bibnamefont
		{Winter}}\ and\ \bibinfo {author} {\bibfnamefont {M.}\ \bibnamefont {Lewenstein}},\
}\bibfield  {title} {\enquote {\bibinfo {title} {Entanglement and coherence in quantum state merging},}\ }\href {\doibase
	10.1103/PhysRevLett.116.240405} {\bibfield  {journal} {\bibinfo  {journal}
		{Phys. Rev. Lett.}\ }\textbf {\bibinfo {volume} {116}},\ \bibinfo {pages}
	{240405} (\bibinfo {year} {2016})}\BibitemShut {NoStop}
\bibitem [{\citenamefont {Santos2019}\ \emph {et~al.}(2019)\citenamefont {Santos},
	\citenamefont {Céleri}, \citenamefont {Landi} \ and\ \citenamefont {Paternostro}}]{Santos2019}
\BibitemOpen
\bibfield  {author} {\bibinfo {author} {\bibfnamefont {J. P.}\ \bibnamefont
		{Santos}}, \bibinfo {author} {\bibfnamefont {L. C.}~\bibnamefont
		{Céleri}}, \bibinfo {author} {\bibfnamefont {G. T.}~\bibnamefont
		{Landi}}\ and\ \bibinfo {author} {\bibfnamefont {M.}\ \bibnamefont {Paternostro}},\
}\bibfield  {title} {\enquote {\bibinfo {title} {The role of quantum coherence in non-equilibrium entropy production},}\ }\href {\doibase
	10.1038/s41534-019-0138-y} {\bibfield  {journal} {\bibinfo  {journal}
		{npj. Quantum. Inf.}\ }\textbf {\bibinfo {volume} {5}},\ \bibinfo {pages}
	{1-7} (\bibinfo {year} {2019})}\BibitemShut {NoStop}
\bibitem [{\citenamefont {Narasimhachar2015}\ \emph {et~al.}(2015)\citenamefont {Narasimhachar} \ and\ \citenamefont {Gour}}]{Narasimhachar2015}
\BibitemOpen
\bibfield  {author} {\bibinfo {author} {\bibfnamefont {V.}\ \bibnamefont
		{Narasimhachar}}\ and\ \bibinfo {author} {\bibfnamefont {G.}\ \bibnamefont {Gour}},\
}\bibfield  {title} {\enquote {\bibinfo {title} {Low-temperature thermodynamics with quantum coherence},}\ }\href {\doibase
	10.1038/ncomms8689} {\bibfield  {journal} {\bibinfo  {journal} {Nat. Commun.}\ }\textbf {\bibinfo {volume} {6}},\ \bibinfo {pages}
	{7689} (\bibinfo {year} {2015})}\BibitemShut {NoStop}
\bibitem [{\citenamefont {Carollo2003}\ \emph {et~al.}(2003)\citenamefont {Carollo},
	\citenamefont {Santos} \ and\ \citenamefont {Vedral}}]{Carollo2003}
\BibitemOpen
\bibfield  {author} {\bibinfo {author} {\bibfnamefont {A.}\ \bibnamefont
		{Carollo}}, \bibinfo {author} {\bibfnamefont {M. F.}~\bibnamefont
		{Santos}}\ and\ \bibinfo {author} {\bibfnamefont {V.}\ \bibnamefont {Vedral}},\
}\bibfield  {title} {\enquote {\bibinfo {title} {Berry’s phase in cavity QED: Proposal for observing an effect of field quantization},}\ }\href {\doibase
	10.1103/PhysRevA.67.063804} {\bibfield  {journal} {\bibinfo  {journal}
		{Phys. Rev. A.}\ }\textbf {\bibinfo {volume} {67}},\ \bibinfo {pages}
	{063804} (\bibinfo {year} {2003})}\BibitemShut {NoStop}
\bibitem [{\citenamefont {Salah2018}\ \emph {et~al.}(2018)\citenamefont {Salah},
	\citenamefont {Abdel-Rady}, \citenamefont {Osman} \ and\ \citenamefont {Hassan}}]{Salah2018}
\BibitemOpen
\bibfield  {author} {\bibinfo {author} {\bibfnamefont {A.}\ \bibnamefont
		{Salah}}, \bibinfo {author} {\bibfnamefont {A. S.}~\bibnamefont
		{Abdel-Rady}}, \bibinfo {author} {\bibfnamefont {A. N. A.}~\bibnamefont
		{Osman}}\ and\ \bibinfo {author} {\bibfnamefont {S. S.}\ \bibnamefont {Hassan}},\
}\bibfield  {title} {\enquote {\bibinfo {title} {Enhancing quantum phase transitions in the critical point of Extended TC-Dicke model via Stark effect},}\ }\href {\doibase
	10.1038/s41598-018-29902-9} {\bibfield  {journal} {\bibinfo  {journal}
		{Sci. Rep.}\ }\textbf {\bibinfo {volume} {8}},\ \bibinfo {pages}
	{11633} (\bibinfo {year} {2018})}\BibitemShut {NoStop}
\bibitem [{\citenamefont {Abdel-Rady2017}\ \emph {et~al.}(2017)\citenamefont {Abdel-Rady},
	\citenamefont {Hassan}, \citenamefont {Osman} \ and\ \citenamefont {Salah}}]{Abdel-Rady2017}
\BibitemOpen
\bibfield  {author} {\bibinfo {author} {\bibfnamefont {A.S.}\ \bibnamefont
		{Abdel-Rady}}, \bibinfo {author} {\bibfnamefont {S. S.}~\bibnamefont
		{Hassan}}, \bibinfo {author} {\bibfnamefont {A. N. A.}~\bibnamefont
		{Osman}}\ and\ \bibinfo {author} {\bibfnamefont {A.}\ \bibnamefont {Salah}},\
}\bibfield  {title} {\enquote {\bibinfo {title} {Quantum phase transition and Berry phase of the Dicke model in the presence of the Stark-shift},}\ }\href {\doibase
	10.1142/S0217979217500916} {\bibfield  {journal} {\bibinfo  {journal}
		{Int. J. Mod. Phys. B.}\ }\textbf {\bibinfo {volume} {31}},\ \bibinfo {pages}
	{1750091} (\bibinfo {year} {2017})}\BibitemShut {NoStop}
\bibitem [{\citenamefont {Alsing1987}\ \emph {et~al.}(1987)\citenamefont {Alsing} \ and\ \citenamefont {Zubairy}}]{Alsing1987}
\BibitemOpen
\bibfield  {author} {\bibinfo {author} {\bibfnamefont {P.}\ \bibnamefont
		{Alsing}} \ and\ \bibinfo {author} {\bibfnamefont {M. S.}\ \bibnamefont {Zubairy}},\
}\bibfield  {title} {\enquote {\bibinfo {title} {Collapse and revivals in a two-photon absorption process},}\ }\href {\doibase
	10.1364/JOSAB.4.000177} {\bibfield  {journal} {\bibinfo  {journal}
		{J. Opt. Soc. Am. B.}\ }\textbf {\bibinfo {volume} {4}},\ \bibinfo {pages}
	{177-184} (\bibinfo {year} {1987})}\BibitemShut {NoStop}
\bibitem [{\citenamefont {Puri1988}\ \emph {et~al.}(1988)\citenamefont {Alsing} \ and\ \citenamefont {Bullough}}]{Puri1988}
\BibitemOpen
\bibfield  {author} {\bibinfo {author} {\bibfnamefont {R. R.}\ \bibnamefont
		{Puri}} \ and\ \bibinfo {author} {\bibfnamefont {R. K.}\ \bibnamefont {Bullough}},\
}\bibfield  {title} {\enquote {\bibinfo {title} {Quantum electrodynamics of an atom making two-photon transitions in an ideal cavity},}\ }\href {\doibase
	10.1364/JOSAB.5.002021} {\bibfield  {journal} {\bibinfo  {journal}
		{J. Opt. Soc. Am. B.}\ }\textbf {\bibinfo {volume} {5}},\ \bibinfo {pages}
	{2021-2028} (\bibinfo {year} {1988})}\BibitemShut {NoStop}
\bibitem [{\citenamefont {Ashraf1990}\ \emph {et~al.}(1990)\citenamefont {Ashraf} \ and\ \citenamefont {Zubairy}}]{Ashraf1990}
\BibitemOpen
\bibfield  {author} {\bibinfo {author} {\bibfnamefont {I.}\ \bibnamefont
		{Ashraf}} \ and\ \bibinfo {author} {\bibfnamefont {M. S.}\ \bibnamefont {Zubairy}},\
}\bibfield  {title} {\enquote {\bibinfo {title} {Photon statistics of the two-photon micromaser},}\ }\href {\doibase
	10.1016/0030-4018(90)90466-7} {\bibfield  {journal} {\bibinfo  {journal}
		{Opt. Commun.}\ }\textbf {\bibinfo {volume} {77}},\ \bibinfo {pages}
	{85-90} (\bibinfo {year} {1990})}\BibitemShut {NoStop}
\bibitem [{\citenamefont {Gou1990}\ \emph {et~al.}(1990)\citenamefont {Gou}}]{Gou1990}
\BibitemOpen
\bibfield  {author} {\bibinfo {author} {\bibfnamefont {S.-C.}\ \bibnamefont
		{Gou}},\
}\bibfield  {title} {\enquote {\bibinfo {title} {Dynamics of the two-mode Jaynes-Cummings model modified by Stark shifts.},}\ }\href {\doibase
	10.1016/0375-9601(90)90635-2} {\bibfield  {journal} {\bibinfo  {journal}
		{Phys. Lett. A.}\ }\textbf {\bibinfo {volume} {147}},\ \bibinfo {pages}
	{218-222} (\bibinfo {year} {1990})}\BibitemShut {NoStop}
\bibitem [{\citenamefont {Yu2016}\ \emph {et~al.}(2016)\citenamefont {Yu},
	\citenamefont {Zhang} \citenamefont {Xu} \ and\ \citenamefont {Tong}}]{Yu2016}
\BibitemOpen
\bibfield  {author} {\bibinfo {author} {\bibfnamefont {X.-D.}\ \bibnamefont
		{Yu}}, \bibinfo {author} {\bibfnamefont {D.-J.}~\bibnamefont
		{Zhang}}, \bibinfo {author} {\bibfnamefont {G. F.}~\bibnamefont
		{Xu}}\ and\ \bibinfo {author} {\bibfnamefont {D. M.}\ \bibnamefont {Tong}},\
}\bibfield  {title} {\enquote {\bibinfo {title} {Alternative framework for quantifying coherence},}\ }\href {\doibase 10.1103/PhysRevA.94.060302} {\bibfield  {journal} {\bibinfo  {journal}
		{Phys. Rev. A.}\ }\textbf {\bibinfo {volume} {94}},\ \bibinfo {pages}
	{060302} (\bibinfo {year} {2016})}\BibitemShut {NoStop}
\bibitem [{\citenamefont {Winter2016}\ \emph {et~al.}(2016)\citenamefont {Winter} \ and\ \citenamefont {Yang}}]{Winter2016}
\BibitemOpen
\bibfield  {author} {\bibinfo {author} {\bibfnamefont {A.}\ \bibnamefont
		{Winter}} \bibinfo {author} \ and\ \bibinfo {author} {\bibfnamefont {D.}\ \bibnamefont {Yang}},\
}\bibfield  {title} {\enquote {\bibinfo {title} {Operational Resource Theory of Coherence},}\ }\href {\doibase 10.1103/PhysRevLett.116.120404} {\bibfield  {journal} {\bibinfo  {journal}
		{Phys. Rev. Lett.}\ }\textbf {\bibinfo {volume} {116}},\ \bibinfo {pages}
	{120404} (\bibinfo {year} {2016})}\BibitemShut {NoStop}
\bibitem [{\citenamefont {Streltsov2015}\ \emph {et~al.}(2015)\citenamefont {Streltsov},
	\citenamefont {Li}, \citenamefont {Nie}, \citenamefont {Li} \ and\ \citenamefont {Adesso}}]{Streltsov2015}
\BibitemOpen
\bibfield  {author} {\bibinfo {author} {\bibfnamefont {A.}\ \bibnamefont
		{Streltsov}}, \bibinfo {author} {\bibfnamefont {U.}~\bibnamefont
		{Singh}}, \bibinfo {author} {\bibfnamefont {H. S.}~\bibnamefont
		{Dhar}}, \bibinfo {author} {\bibfnamefont {M. N.}~\bibnamefont
		{Bera}} \ and\ \bibinfo {author} {\bibfnamefont {G.}\ \bibnamefont {Adesso}},\
}\bibfield  {title} {\enquote {\bibinfo {title} {Measuring quantum coherence with entanglement},}\ }\href {\doibase
	10.1103/PhysRevLett.115.020403} {\bibfield  {journal} {\bibinfo  {journal}
		{Phys. Rev. Lett.}\ }\textbf {\bibinfo {volume} {115}},\ \bibinfo {pages}
	{020403} (\bibinfo {year} {2015})}\BibitemShut {NoStop}
\bibitem [{\citenamefont {Yuan2015}\ \emph {et~al.}(2015)\citenamefont {Yuan}, \citenamefont {Zhou}, \citenamefont {Cao} \ and\ \citenamefont {Ma}}]{Yuan2015}
\BibitemOpen
\bibfield  {author} {\bibinfo {author} {\bibfnamefont {X.}\ \bibnamefont
		{Yuan}}, \bibinfo {author} {\bibfnamefont {H.}~\bibnamefont
		{Zhou}}, \bibinfo {author} {\bibfnamefont {Z.}~\bibnamefont
		{Cao}} \ and\ \bibinfo {author} {\bibfnamefont {X.}~\bibnamefont
		{Ma}},\
}\bibfield  {title} {\enquote {\bibinfo {title} {Intrinsic randomness as a measure of quantum coherence},}\ }\href {\doibase
	10.1103/PhysRevA.92.022124} {\bibfield  {journal} {\bibinfo  {journal}
		{Phys. Rev. A.}\ }\textbf {\bibinfo {volume} {92}},\ \bibinfo {pages}
	{022124} (\bibinfo {year} {2015})}\BibitemShut {NoStop}
\bibitem [{\citenamefont {Qi2017}\ \emph {et~al.}(2017)\citenamefont {Qi}, \citenamefont {Gao} \ and\ \citenamefont {Yan}}]{Qi2017}
\BibitemOpen
\bibfield  {author} {\bibinfo {author} {\bibfnamefont {X.}\ \bibnamefont
		{Qi}}, \bibinfo {author} {\bibfnamefont {T.}~\bibnamefont
		{Gao}} \ and\ \bibinfo {author} {\bibfnamefont {F.}~\bibnamefont
		{Yan}},\
}\bibfield  {title} {\enquote {\bibinfo {title} {Measuring coherence with entanglement concurrence},}\ }\href {\doibase
	10.1088/1751-8121/aa7638} {\bibfield  {journal} {\bibinfo  {journal}
		{J. Phys. A: Mathematical and Theoretical.}\ }\textbf {\bibinfo {volume} {50}},\ \bibinfo {pages}
	{285301} (\bibinfo {year} {2017})}\BibitemShut {NoStop}
\bibitem [{\citenamefont {Yu2017}\ \emph {}(2017)\citenamefont {Yu}}]{Yu2017}
\BibitemOpen
\bibfield  {author} {\bibinfo {author} {\bibfnamefont {C.-S.}\ \bibnamefont
		{Yu}},\
}\bibfield  {title} {\enquote {\bibinfo {title} {Quantum coherence via skew information and its polygamy},}\ }\href {\doibase
10.1103/PhysRevA.95.042337} {\bibfield  {journal} {\bibinfo  {journal}
	{Phys. Rev. A.}\ }\textbf {\bibinfo {volume} {95}},\ \bibinfo {pages}
{042337} (\bibinfo {year} {2017})}\BibitemShut {NoStop}
\bibitem [{\citenamefont {Rastegin2016}\ \emph {et~al.}(2016)\citenamefont {Rastegin} }]{Rastegin2016}
\BibitemOpen
\bibfield  {author} {\bibinfo {author} {\bibfnamefont {A. E.}\ \bibnamefont {Rastegin}},\
}\bibfield  {title} {\enquote {\bibinfo {title} {Quantum-coherence quantifiers based on the Tsallis relative $\alpha$ entropies},}\ }\href {\doibase
	10.1103/PhysRevA.93.032136} {\bibfield  {journal} {\bibinfo  {journal}
		{Phys. Rev. A.}\ }\textbf {\bibinfo {volume} {93}},\ \bibinfo {pages}
	{032136} (\bibinfo {year} {2016})}\BibitemShut {NoStop}
\bibitem [{\citenamefont {Zhao2018}\ \emph {et~al.}(2018)\citenamefont {Zhao} \ and\ \citenamefont {Yu}}]{Zhao2018}
\BibitemOpen
\bibfield  {author} {\bibinfo {author} {\bibfnamefont {H.}\ \bibnamefont
		{Zhao}}\ and\ \bibinfo {author} {\bibfnamefont {C. S.}\ \bibnamefont {Yu}},\
}\bibfield  {title} {\enquote {\bibinfo {title} {Coherence measure in terms of the Tsallis relative $\alpha$ entropy},}\ }\href {\doibase
	10.1103/PhysRevLett.88.017901} {\bibfield  {journal} {\bibinfo  {journal}
		{Sci. Rep.}\ }\textbf {\bibinfo {volume} {8}},\ \bibinfo {pages}
	{299} (\bibinfo {year} {2018})}\BibitemShut {NoStop}
\bibitem [{\citenamefont {Liu2017}\ \emph {et~al.}(2017)\citenamefont {Liu}, \citenamefont {Zhang}, \citenamefont {Yu}, \citenamefont {Ding} \ and\ \citenamefont {Liu}}]{Liu2017}
\BibitemOpen
\bibfield  {author} {\bibinfo {author} {\bibfnamefont {C. L.}\ \bibnamefont
		{Liu}}, \bibinfo {author} {\bibfnamefont {D.-J.}~\bibnamefont
		{Zhang}}, \bibinfo {author} {\bibfnamefont {X.-D.}~\bibnamefont
		{Yu}}, \bibinfo {author} {\bibfnamefont {Q.-M.}~\bibnamefont
		{Ding}} \ and\ \bibinfo {author} {\bibfnamefont {L.}\ \bibnamefont {Liu}},\
}\bibfield  {title} {\enquote {\bibinfo {title} {A new coherence measure based on fidelity},}\ }\href {\doibase
	10.1007/s11128-017-1650-7} {\bibfield  {journal} {\bibinfo  {journal}
		{Quantum Inf Process.}\ }\textbf {\bibinfo {volume} {16}},\ \bibinfo {pages}
	{198} (\bibinfo {year} {2017})}\BibitemShut {NoStop}
\bibitem [{\citenamefont {Xu2016}\ \emph {et~al.}(2016)}]{Xu2016}
\BibitemOpen
\bibfield  {author} {\bibinfo {author} {\bibfnamefont {J.}\ \bibnamefont
		{Xu}},\
}\bibfield  {title} {\enquote {\bibinfo {title} {Quantifying coherence of Gaussian states},}\ }\href {\doibase
	10.1103/PhysRevA.93.032111} {\bibfield  {journal} {\bibinfo  {journal}
		{Phys. Rev. A.}\ }\textbf {\bibinfo {volume} {93}},\ \bibinfo {pages}
	{032111} (\bibinfo {year} {2016})}\BibitemShut {NoStop}
\bibitem [{\citenamefont {Radhakrishnan2016}\ \emph {et~al.}(2016)\citenamefont {Radhakrishnan},
	\citenamefont {Parthasarathy}, \citenamefont {Jambulingam} \ and\ \citenamefont {Byrnes}}]{Radhakrishnan2016}
\BibitemOpen
\bibfield  {author} {\bibinfo {author} {\bibfnamefont {C.}\ \bibnamefont
		{Radhakrishnan}}, \bibinfo {author} {\bibfnamefont {M.}~\bibnamefont
		{Parthasarathy}}, \bibinfo {author} {\bibfnamefont {S.}~\bibnamefont
		{Jambulingam}}\ and\ \bibinfo {author} {\bibfnamefont {T.}\ \bibnamefont {Byrnes}},\
}\bibfield  {title} {\enquote {\bibinfo {title} {Distribution of quantum coherence in multipartite systems},}\ }\href {\doibase
	10.1103/PhysRevLett.116.150504} {\bibfield  {journal} {\bibinfo  {journal}
		{Phys. Rev. Lett.}\ }\textbf {\bibinfo {volume} {116}},\ \bibinfo {pages}
	{150504} (\bibinfo {year} {2016})}\BibitemShut {NoStop}
\bibitem [{\citenamefont {Lin1991}\ \emph {et~al.}(1991)\citenamefont {Åberg} }]{Lin1991}
\BibitemOpen
\bibfield  {author} {\bibinfo {author} {\bibfnamefont {J.}\ \bibnamefont {Lin}},\
}\bibfield  {title} {\enquote {\bibinfo {title} {Divergence measures based on the Shannon entropy},}\ }\href {\doibase
	10.1109/18.61115} {\bibfield  {journal} {\bibinfo  {journal}
		{IEEE Transactions on Information theory.}\ }\textbf {\bibinfo {volume} {37}},\ \bibinfo {pages}
	{145-151} (\bibinfo {year} {1991})}\BibitemShut {NoStop}
\bibitem [{\citenamefont {Briet2009}\ \emph {et~al.}(2009)\citenamefont {Briët} \ and\ \citenamefont {Harremoës}}]{Briet2009}
\BibitemOpen
\bibfield  {author} {\bibinfo {author} {\bibfnamefont {J.}\ \bibnamefont
		{Briët}} \ and\ \bibinfo {author} {\bibfnamefont {P.}\ \bibnamefont {Harremoës}},\
}\bibfield  {title} {\enquote {\bibinfo {title} {Properties of classical and quantum Jensen-Shannon divergence},}\ }\href {\doibase
	10.1103/PhysRevA.79.052311} {\bibfield  {journal} {\bibinfo  {journal}
		{Phys. Rev. A.}\ }\textbf {\bibinfo {volume} {79}},\ \bibinfo {pages}
	{052311} (\bibinfo {year} {2009})}\BibitemShut {NoStop}\bibitem [{\citenamefont {Majtey2005}\ \emph {et~al.}(2005)\citenamefont {Majtey}, \citenamefont {Lamberti} \ and\ \citenamefont {Prato}}]{Majtey2005}
\BibitemOpen
\bibfield  {author} {\bibinfo {author} {\bibfnamefont {A.P.}\ \bibnamefont
		{Majtey}}, \bibinfo {author} {\bibfnamefont {P.W.}\ \bibnamefont {Lamberti}}, \ and\ \bibinfo {author} {\bibfnamefont {D.P.}\ \bibnamefont {Prato}},\
}\bibfield  {title} {\enquote {\bibinfo {title} {Jensen-Shannon divergence as a measure of distinguishability between mixed quantum states},}\ }\href {\doibase
	10.1103/PhysRevA.72.052310} {\bibfield  {journal} {\bibinfo  {journal}
		{Phys. Rev. A.}\ }\textbf {\bibinfo {volume} {72}},\ \bibinfo {pages}
	{052310} (\bibinfo {year} {2005})}\BibitemShut {NoStop}
\bibitem [{\citenamefont {Lamberti2008}\ \emph {et~al.}(2008)\citenamefont {Lamberti}, \citenamefont {Majtey},  {Borras}, \citenamefont {Casas} \ and\ \citenamefont {Plastino}}]{Lamberti2008}
\BibitemOpen
\bibfield  {author} {\bibinfo {author} {\bibfnamefont {P.W.}\ \bibnamefont
		{Lamberti}}, \bibinfo {author} {\bibfnamefont {A.P.}\ \bibnamefont {Majtey}}, \bibinfo {author} {\bibfnamefont {A.}\ \bibnamefont {Borras}}, \bibinfo {author} {\bibfnamefont {M.}\ \bibnamefont {Casas}}, \ and\ \bibinfo {author} {\bibfnamefont {A.}\ \bibnamefont {Plastino}},\
}\bibfield  {title} {\enquote {\bibinfo {title} {Metric character of the quantum Jensen-Shannon divergence},}\ }\href {\doibase
	10.1103/PhysRevA.77.052311} {\bibfield  {journal} {\bibinfo  {journal}
		{Phys. Rev. A.}\ }\textbf {\bibinfo {volume} {77}},\ \bibinfo {pages}
	{052311} (\bibinfo {year} {2008})}\BibitemShut {NoStop}
\bibitem [{\citenamefont {Dakic2010}\ \emph {et~al.}(2010)\citenamefont {Dakic},
	\citenamefont {Vedral} \ and\ \citenamefont {Brukner}}]{Dakic2010}
\BibitemOpen
\bibfield  {author} {\bibinfo {author} {\bibfnamefont {B.}\ \bibnamefont
		{Dakic}}, \bibinfo {author} {\bibfnamefont {V.}~\bibnamefont
		{Vedral}} \ and\ \bibinfo {author} {\bibfnamefont {C.}\ \bibnamefont {Brukner}},\
}\bibfield  {title} {\enquote {\bibinfo {title} {Necessary and sufficient condition for nonzero quantum discord},}\ }\href {\doibase
	10.1103/PhysRevLett.105.190502} {\bibfield  {journal} {\bibinfo  {journal}
		{Phys. Rev. Lett.}\ }\textbf {\bibinfo {volume} {105}},\ \bibinfo {pages}
	{190502} (\bibinfo {year} {2010})}\BibitemShut {NoStop}
\bibitem [{\citenamefont {Girolami2011}\ \emph {et~al.}(2011)\citenamefont {Girolami} \ and\ \citenamefont {Adesso}}]{Girolami2011}
\BibitemOpen
\bibfield  {author} {\bibinfo {author} {\bibfnamefont {D.}\ \bibnamefont
		{Girolami}} \ and\ \bibinfo {author} {\bibfnamefont {G.}\ \bibnamefont {Adesso}},\
}\bibfield  {title} {\enquote {\bibinfo {title} {Quantum discord for general two-qubit states: analytical progress},}\ }\href {\doibase
	10.1103/PhysRevA.83.052108} {\bibfield  {journal} {\bibinfo  {journal}
		{Phys. Rev. A.}\ }\textbf {\bibinfo {volume} {83}},\ \bibinfo {pages}
	{052108} (\bibinfo {year} {2011})}\BibitemShut {NoStop}
\bibitem [{\citenamefont {Wang2010}\ \emph {et~al.}(2010)\citenamefont {Wang},
  \citenamefont {Li}, \citenamefont {Nie} \ and\ \citenamefont {Li}}]{Wang2010}
  \BibitemOpen
  \bibfield  {author} {\bibinfo {author} {\bibfnamefont {C.-Z.}\ \bibnamefont
  {Wang}}, \bibinfo {author} {\bibfnamefont {C.-X.}~\bibnamefont
  {Li}}, \bibinfo {author} {\bibfnamefont {L.-Y.}~\bibnamefont
  {Nie}}\ and\ \bibinfo {author} {\bibfnamefont {J.-F.}\ \bibnamefont {Li}},\
  }\bibfield  {title} {\enquote {\bibinfo {title} {Classical correlation and quantum discord mediated by cavity in two coupled qubits},}\ }\href {\doibase
  10.1088/0953-4075/44/1/015503} {\bibfield  {journal} {\bibinfo  {journal}
  {J. Phys. B: At.﻿Mol. Opt. Phys.}\ }\textbf {\bibinfo {volume} {44}},\ \bibinfo {pages}
  {015503} (\bibinfo {year} {2010})}\BibitemShut {NoStop}
\bibitem [{\citenamefont {Yao2015}\ \emph {et~al.}(2015)\citenamefont {Yao},
	\citenamefont {Xiao}, \citenamefont {Ge} \ and\ \citenamefont {Sun}}]{Yao2015}
\BibitemOpen
\bibfield  {author} {\bibinfo {author} {\bibfnamefont {Y.}\ \bibnamefont
		{Yao}}, \bibinfo {author} {\bibfnamefont {X.}~\bibnamefont
		{Xiao}}, \bibinfo {author} {\bibfnamefont {L.}~\bibnamefont
		{Ge}}\ and\ \bibinfo {author} {\bibfnamefont {C. P.}\ \bibnamefont {Sun}},\
}\bibfield  {title} {\enquote {\bibinfo {title} {Quantum coherence in multipartite systems},}\ }\href {\doibase
	10.1103/PhysRevA.92.022112} {\bibfield  {journal} {\bibinfo  {journal}
		{Phys. Rev. A.}\ }\textbf {\bibinfo {volume} {92}},\ \bibinfo {pages}
	{022112} (\bibinfo {year} {2015})}\BibitemShut {NoStop}
\bibitem [{\citenamefont {Huang2019}\ \emph {et~al.}(2019)\citenamefont {Huang}}]{Huang2019}
\BibitemOpen
\bibfield  {author} {\bibinfo {author} {\bibfnamefont {Z.}\ \bibnamefont
		{Huang}},\
}\bibfield  {title} {\enquote {\bibinfo {title} {Behaviors of quantum correlation for atoms coupled with fluctuating electromagnetic field with a perfectly reflecting boundary},}\ }\href {\doibase
	10.1007/s11128-019-2268-8} {\bibfield  {journal} {\bibinfo  {journal}
		{Quantum Inf. Process.}\ }\textbf {\bibinfo {volume} {18}},\ \bibinfo {pages}
	{149} (\bibinfo {year} {2019})}\BibitemShut {NoStop}
\bibitem [{\citenamefont {Cheng2018}\ \emph {et~al.}(2018)\citenamefont {Cheng}, \citenamefont {Yu} \ and\ \citenamefont {Hu}}]{Cheng2018}
\BibitemOpen
\bibfield  {author} {\bibinfo {author} {\bibfnamefont {S.}\ \bibnamefont
		{Cheng}}, \bibinfo {author} {\bibfnamefont {H.}~\bibnamefont
		{Yu}}\ and\ \bibinfo {author} {\bibfnamefont {J.}\ \bibnamefont {Hu}},\
}\bibfield  {title} {\enquote {\bibinfo {title} {Entanglement dynamics for uniformly accelerated two-level atoms in the presence of a reflecting boundary},}\ }\href {\doibase
	10.1103/PhysRevD.98.025001} {\bibfield  {journal} {\bibinfo  {journal}
		{Phys. Rev. D.}\ }\textbf {\bibinfo {volume} {98}},\ \bibinfo {pages}
	{025001} (\bibinfo {year} {2018})}\BibitemShut {NoStop}
\end{thebibliography}


\end{document}